\documentclass{SciPost}
\usepackage{comment}
\usepackage{caption}
\usepackage{wrapfig}

\hypersetup{
    colorlinks,
    linkcolor={red!50!black},
    citecolor={blue!50!black},
    urlcolor={blue!80!black}
}

\usepackage[bitstream-charter]{mathdesign}
\DeclareSymbolFont{usualmathcal}{OMS}{cmsy}{m}{n}
\DeclareSymbolFontAlphabet{\mathcal}{usualmathcal}

\fancypagestyle{SPstyle}{
\fancyhf{}
\lhead{\colorbox{scipostblue}{\bf \color{white} ~SciPost Physics Community Reports }}
\rhead{{\bf \color{scipostdeepblue} ~Submission }}

\fancyfoot[C]{\textbf{\thepage}}
}

\begin{document}

\pagestyle{SPstyle}
\begin{center}{\large \textbf{
Are We Ready for AI-Driven Discovery?\\
AI Verification Before the Next Fundamental Physics Breakthrough\\
}
}
\end{center}

\begin{center}\textbf{
Gaia Grosso\textsuperscript{$\star,\dagger$,1,2,3},
Vinicius Mikuni\textsuperscript{$\star,\ddagger$, 4},
Lukas Heinrich\textsuperscript{$\circ$,5,6}
}\end{center}

\begin{center}
{\bf 1} NSF AI Institute for Artificial Intelligence and Fundamental Interactions, Cambridge, MA
\\
{\bf 2} MIT Laboratory for Nuclear Science, Cambridge, MA
\\
{\bf 3} School of Engineering and Applied Sciences, Harvard University, Cambridge, MA
\\
{\bf 4} Nagoya University, Kobayashi-Maskawa Institute, Japan
\\
{\bf 5} Technical University Munich
\\
{\bf 6} Munich Center for Machine Learning (MCML)
\\[\baselineskip]
$\star$ {\small Leading authors}\,,\quad
$\circ$ {\small Advisor}\,,\quad
$\dagger$ \href{mailto:email1}{\small gaiag795@mit.edu}\,,\quad
$\ddagger$ \href{mailto:email2}{\small vmikuni@nagoya-u.jp}
\end{center}

\section*{\color{scipostdeepblue}{Abstract}}
\textbf{\boldmath{
Machine learning (ML) has become integral to fundamental physics, accelerating statistical workflows from data acquisition through inference and hypothesis testing. As ML systems grow increasingly autonomous, ensuring their reliability for discovery claims becomes critical. This review synthesizes the VERaiPHY (Validation \& Evaluation for Robust AI in PHYsics) initiative's frameworks for rigorous ML assessment across particle physics, astrophysics, and cosmology. 
We establish when verification is essential by contextualizing ML within the statistical discovery workflow. We emphasize fundamental limitations: inductive bias is unavoidable, sample complexity bounds learning, and experimental constraints limit discovery. We reflect on physicists' evolving role as both experimental designers and evaluators whose judgments encode scientific rigor into AI systems. Responsible integration requires understanding ML's transformative potential alongside its intrinsic boundaries.
}}

\vspace{\baselineskip}

\vspace{10pt}
\noindent\rule{\textwidth}{1pt}
\tableofcontents
\noindent\rule{\textwidth}{1pt}
\vspace{10pt}


\section{Introduction}\label{sec:intro}
Machine learning (ML) methods have rapidly become integral components of modern scientific workflows, particularly in fundamental physics, where they are now routinely employed for data analysis, simulation, and inference. Beyond improving the precision and accuracy of statistical analyses in searches for new physics, machine learning has recently shown the potential to drastically accelerate the entire scientific workflow, both by learning data representations that expose complex structure in high-dimensional data, and through emerging agentic capabilities that automate higher-level tasks such as experimental design, simulation steering, and iterative inference.
At the same time, the increasing autonomy and complexity of these methods raise pressing questions regarding their scientific reliability, interpretability, and trustworthiness. 

\paragraph{Why verification?} 
A central challenge in fundamental physics is that the discoveries of new laws of Nature emerge through indirect statistical inference rather than direct observation. 
We do not observe quarks, Higgs bosons, or dark matter particles directly. Instead, these phenomena must be inferred from their effects on observables—energy deposits in detectors, angular distributions of decay products, correlations in cosmological fields—that are themselves noisy, incomplete, and contaminated by backgrounds. 
Unlike many scientific fields where signal can be isolated, fundamental physics searches must disentangle rare signal processes from overwhelming backgrounds within high-dimensional, stochastic measurements. Even if one had access to the most powerful discrimination tools achievable, the probabilistic nature of quantum mechanics, which governs particle physics, and the missing information from the universe evolution would still require a robust statistical description of nature. 
Discovery therefore manifests not as a single unambiguous event, but as a statistically significant deviation from the probability distributions predicted by our current best models. The reconciliation of empirical ML success with the stringent requirements of this statistical inference workflow is therefore central to scientific ML in this field.

Classical statistical algorithms often operate in low dimensional spaces and are typically designed with explicit guarantees, often asymptotic properties such as consistency, lack of bias, or nominal coverage. In contrast, many contemporary ML algorithms are deployed as high-dimensional ``black boxes'' whose impressive empirical performance is not accompanied by comparably stronger theoretical or statistical assurances. For instance, a generative model may produce realistic-looking simulated energy deposits on a detector, achieving high fidelity on visual inspection of a single outcome, yet without verification at the distribution level we cannot know whether it faithfully represents the statistical properties of the real generative process. This gap between the true data-generating process and a miscalibrated surrogate model carries concrete scientific risk: analyses built on such surrogates may yield false discoveries, or conversely, obscure genuine signals — undermining the very statistical claims the workflow was designed to support.
Similarly, a classifier may achieve excellent signal-background discrimination, but if it learns to exploit detector artifacts (such as poorly modeled pileup-dependent response or calibration drift correlated with signal rate) rather than desirable physics properties, using its outputs without proper corrections in hypothesis tests will introduce unmodeled systematic biases that corrupt statistical inference.
As a result, when ML tools are used for statistical modeling, inference, or hypothesis testing, it is often unclear whether they satisfy the properties required for scientifically valid deployment—including calibration, robustness to distributional shift, and well-defined error control. Closing this gap, rather than accepting it as an unavoidable trade-off between performance and rigor, is essential for the responsible integration of machine learning into fundamental physics. 


\paragraph{From guarantees a priori to verification a posteriori.}
The verification challenge is particularly relevant for modern ML development. As emphasized in a recent keynote by Kyunghyun Cho at the 2025 NeurIPS conference~\cite{cho2025neurips}, the scale and complexity of modern AI models forces a fundamental shift in how we ensure statistical validity: from enforcing guarantees \textit{a priori} through careful algorithm design, to assessing them \textit{a posteriori} through systematic verification after training.
As models and data grow to billions of parameters, trained on heterogeneous data through stochastic optimization with emergent behaviors, directly imposing statistical properties by construction becomes increasingly impractical. Instead of training specialized models for each individual task, a new paradigm shift is to train a foundation model~\cite{bommasani2021opportunities}, or a large but general model capable of leveraging the data structure and to be efficiently adapted to new scientific tasks. We cannot feasibly prove that a foundation model, fine-tuned for particle physics, will maintain nominal performance across all relevant distributional shifts, nor can we guarantee that a diffusion model~\cite{yang2023diffusion} for fast simulation will preserve all physically relevant correlations. 
Verification after training becomes essential.
Verification protocols provide systematic procedures to evaluate whether a trained model satisfies the assumptions required for a given scientific task, to quantify the regimes in which it can be trusted, and to delineate its range of safe applicability. In this sense, verification plays a critical role in bridging the gap between the empirical success of machine learning methods and the standards of reliability and interpretability demanded in scientific settings. 

\paragraph{About this article.} This article contributes to VERaiPHY (Validation \& Evaluation for Robust AI in PHYsics), a PHYSTAT review series establishing verification and validation standards for machine learning across particle physics, astrophysics, and cosmology. This article serves three primary functions within the VERaiPHY framework: contextualize, inspire, reflect. 
We start in Section~\ref{sec:workflow} contextualizing the role of machine learning within the end-to-end statistical workflow of fundamental physics discovery. We identify the computational and methodological bottlenecks at each stage—data acquisition, summarization, modeling, simulation, inference—and connect them to the ML methods being applied to overcome these limitations. 
We also identify new emerging research directions in the intersection of AI and statistical analysis. 
In Section~\ref{sec:whenwhy}, we address foundational questions that frequently arise in applications of ML to statistical analysis: when is it acceptable for a model to be ``wrong''? When does verification become critical? When do we need epistemic uncertainties? When are interpretability and explainability essential versus optional? We argue that the answers depend critically on where in the discovery workflow ML is deployed and what role its outputs play in subsequent inference. 

In Section~\ref{sec:limits}, we emphasize the fundamental limitations that ML-assisted statistical analyses are subject to and cannot be overcome by scale or optimization alone. Inductive biases, complexity bounds, experimental constraints. 
We conclude in Section~\ref{sec:conclusions} with summary of principled guidelines to approach ML and its verification for reliable scientific discovery, and a reflection on the role of scientists in an environment of rapidly advancing technologies.

 \section{The impact of AI in fundamental physics discoveries}\label{sec:workflow}

As mentioned in Section~\ref{sec:intro}, discovery in fundamental physics is inseparable from rigorous statistical inference.
The nature of statistical inference varies significantly across the experimental regimes encountered in different subfields.

Collider experiments operate under controlled laboratory settings where the initial conditions (beam energies, collision rates, detector configurations) are determined, and measurements can be repeated many times under nominally identical conditions. In a single second, the Large Hadron Collider produces 40 million proton-proton collisions at precisely controlled energies~\cite{evans2008lhc}. 
Statistical power comes from repetition: by accumulating millions of collision events under controlled conditions, physicists can test whether the observed probability distribution over the experimental observables is consistent with the expected theory predictions or shows deviations indicating new phenomena. This naturally fits frequentist hypothesis testing, where significance is assessed based on how often such an excess or larger would occur by chance if only background processes were present. 

While sharing the repeated-event structure of collider physics, gravitational wave astronomy and rare-event searches (such as those for neutrinos and dark matter) operate in regimes of extreme scarcity. The first gravitational wave detection (GW150914)~\cite{PhysRevLett.116.061102} constituted a discovery from a single event, demonstrating that individual observations can carry sufficient statistical weight when the signal is sufficiently improbable under the null hypothesis. Even with current detector networks, gravitational wave detections occur at rates of tens to hundreds per year—orders of magnitude lower than collider event rates—requiring different statistical frameworks that combine information across sparse catalogs for population-level inference. Similarly, direct dark matter searches may accumulate years of exposure seeking recoil signals at rates potentially below one event per ton-year~\cite{aprile2018dark}, and neutrino astronomy detects rare interactions against overwhelming backgrounds in large-volume detectors.

In contrast, in cosmology we observe only a single realization of the Universe through incomplete and biased measurements of large-scale structure, cosmic microwave background anisotropies, or the distribution of galaxies. Unlike particle physics experiments where repeated events enable frequentist hypothesis testing, cosmology typically proceeds through Bayesian parameter estimation: inferring cosmological parameters, initial conditions, and dynamical laws from spatial correlations in observed fields. Discovery here means ruling out parameter regions or demonstrating that models without specific components (dark matter, dark energy, inflation) cannot explain observations. Note that this mapping reflects convention and natural fit rather than necessity: in principle either framework can be brought to bear on most problems, and analysts sometimes report both as a robustness cross-check — though the frequentist notion of repeated sampling is more natural in some regimes (colliders) than others (a single-realization Universe).

\subsection{The End-to-End Experimental Workflow}
\begin{figure}[t]
    \centering
    \includegraphics[width=0.99\linewidth]{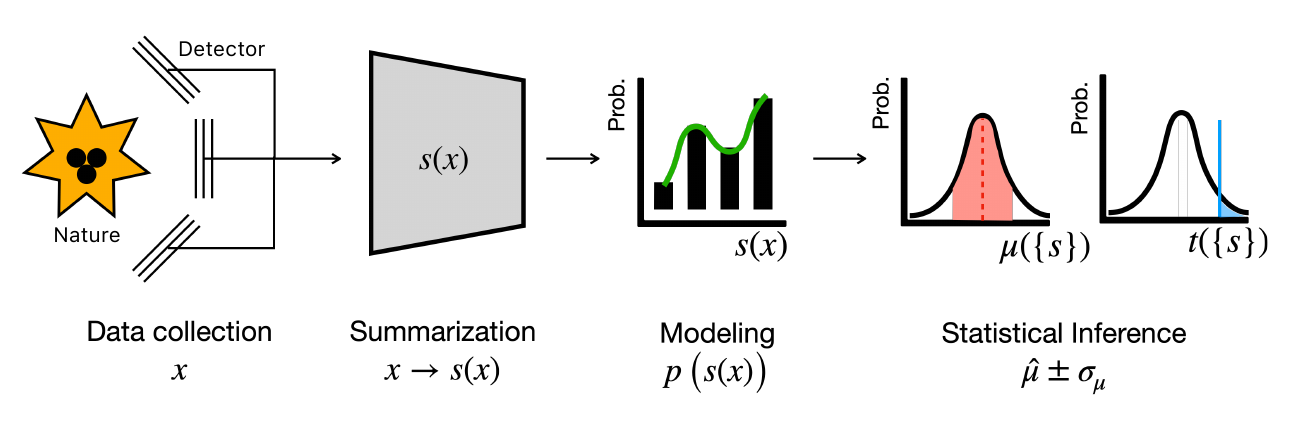}
    \caption{\textbf{The statistical workflow of fundamental physics experiments.} The natural phenomenon is observed using detectors. The collected raw data are transformed into \textit{summary statistics} ($s(x)$), that are then \textit{modeled} and used for \textit{statistical inference} tasks including parameter estimation and statistical tests.}
    \label{fig:workflow}
\end{figure}
Despite differences in observational paradigms, a common statistical workflow underlies discovery across fundamental physics domains. This workflow, visualized in Figure~\ref{fig:workflow}, consists of four interconnected stages:
\begin{enumerate}
    \item \textbf{Data collection} defines what information about physical processes can be accessed. This includes designing detectors or telescopes, choosing which measurements to make, and determining how observations are triggered or scheduled. In colliders, this means deciding detector geometries and trigger thresholds; in cosmology, selecting survey areas, exposure times, and wavelength coverage; in gravitational wave observatories, configuring interferometer sensitivities and data acquisition systems. This stage is largely irreversible: information not captured cannot be recovered in later analyses.
    
    \item \textbf{Summarization} transforms raw experimental signals into meaningful observables, suitable to enhance the signal target. For colliders, this means reconstructing particle properties from detector hits. For cosmological surveys, it involves extracting galaxy catalogs, measuring shapes and redshifts, or constructing maps of cosmic fields. For gravitational wave detectors, it includes identifying candidate events and estimating their time-frequency characteristics. 
    
    \item \textbf{Statistical modeling} constructs probabilistic descriptions of how observables relate to underlying physical parameters. This requires simulating or analytically estimating what data should look like under different theoretical hypotheses, characterizing backgrounds and systematic uncertainties, and building likelihood functions and posterior distributions. The model encodes our understanding of both the physics of interest and the experimental apparatus.
    
    \item \textbf{Statistical inference} extracts physical conclusions by comparing observed data to models. This includes goodness-of-fit tests (does the Standard Model adequately describe the data?), hypothesis testing (is there evidence for new physics beyond the Standard Model? Does inflation fit significantly better than alternatives?), parameter estimation (measuring masses, coupling constants, cosmological parameters), and model selection (which theoretical framework best describes observations?). Throughout, uncertainties must be quantified and propagated to support scientifically valid conclusions.
\end{enumerate}

\paragraph{The central role of simulation.} 
In fundamental physics, simulations play a foundational role across all stages of this workflow, distinguishing this scientific field from most other machine learning application domains. What distinguishes fundamental physics from most other ML application domains is not merely the scale of simulation but the depth of the latent stochastic evolution connecting theory parameters to observations. In many experimental sciences, the relationship between parameters and data is well-approximated by a shallow generative model---effectively $x \sim p(x|\theta) = f_{\mathrm{det}}(\theta) + \varepsilon$, where a deterministic forward model maps parameters to expected outcomes and a single noise term captures measurement uncertainty. In fundamental physics, by contrast, the path from a theoretical parameter $\theta$ to an observation $x$ involves a long cascade of stochastic transitions, each governed by its own complex conditional distribution. In collider physics, this chain runs from hard scattering through parton showering, hadronization, and detector response; in cosmology, from primordial fluctuations through structure formation, baryonic processes, and instrumental systematics; in gravitational wave astronomy, from binary merger dynamics through waveform generation, noise processes, and detector transfer functions. Across all these cases, the likelihood $p(x|\theta)$ is an intractable integral over an enormous latent space with no closed-form expression and no simple noise model. 
\begin{wrapfigure}{r}{0.4\textwidth}
    \centering
    \vspace{-1\baselineskip}
    \includegraphics[width=\linewidth]{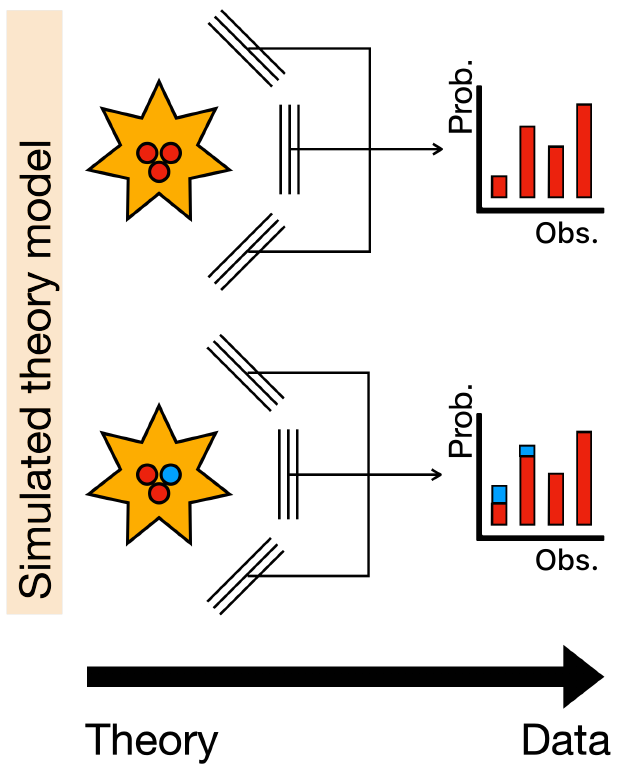}
    \caption{Two different physical phenomena (top row and bottom row) give rise to two different signatures in the observable data representations (on the right side). Simulations allow to probe these properties, ``folding" theories into the observable data representations, to design and validate the statistical workflow before analyzing real data.}
    \label{fig:simulations}
\end{wrapfigure}
This is precisely why simulation is not merely convenient but necessary: it provides the only tractable mechanism for sampling from this deep generative process, and has driven the development of an entire ecosystem of simulation-based inference methods~\cite{cranmer2020frontier} that bypass explicit likelihood evaluation entirely.
Simulations serve as the mechanism through which theoretical predictions are ``folded'' into observable data representations, effectively constructing the null hypothesis against which discoveries are tested. They enable optimization of experimental design by predicting statistical properties under expected conditions, provide labeled datasets for training reconstruction and classification algorithms, and allow systematic validation of analysis procedures under controlled variations. In collider physics, billions of simulated events model both physics processes and detector response; in cosmology, simulation suites spanning parameter spaces enable emulator construction and statistical inference; in gravitational wave astronomy, numerical relativity calculations provides waveform templates for matched 
filtering.

This ability to generate rigorous synthetic experiments—conducting controlled trials of the full analysis chain under precisely known conditions—represents a crucial advantage: we can systematically study how methods perform, we can test statistical properties, and validate procedures before applying them to real data. However, simulation also introduces risks. If simulations do not accurately reflect reality due to modeling approximations, incomplete physics, or miscalibrated detector descriptions, then biases can propagate through the entire analysis chain. This can lead to poor reconstruction, incorrect background estimates, miscalibrated uncertainties, and ultimately false discoveries. Unlike typical ML applications where distribution shift is treated as an unavoidable challenge, fundamental physics experiments invest substantial effort in explicitly modeling domain shifts: characterizing how detector conditions drift, how theoretical predictions vary within their uncertainties, and how systematic effects alter observables. This practice of modeling and propagating systematic uncertainties is essential for maintaining the statistical rigor required for discovery claims.

\subsection{Computational Bottlenecks and ML Solutions}
Machine learning addresses specific computational bottlenecks at each stage of the discovery workflow, enabling scalable, accelerated, and enhanced solutions to new physics discovery. 
We outline in the following the main challenges that ML methods provide new solutions to:

\paragraph{High-rate event selection.} In collider experiments, trigger systems face the daunting task of selecting interesting events from collision rates of 40 MHz while making decisions in microseconds with partially reconstructed information~\cite{khachatryan2017cms,atlas2020operation}. 
In gravitational wave observatories, the challenge is to identify weak, transient signals in non-stationary noise in real-time, across a high-dimensional parameter space, while maintaining strict control of false-alarm rates and latency for multi-messenger follow-up. In dark matter experiments, triggering is less constrained by rate and latency than in collider or gravitational-wave experiments, but becomes critical at ultra-low energy thresholds, where noise, stability, and precise modeling of trigger efficiency directly impact sensitivity and discovery claims~\cite{arnaud2019precision,abdelhameed2019first}.
 
Machine-learning methods are increasingly being explored as solutions to triggering challenges across experiments, offering flexible representations and fast inference in regimes where classical algorithms face scaling limitations. In collider experiments, ML-based triggers enable real-time selection using high-dimensional detector information, improving signal efficiency under strict bandwidth constraints at facilities such as the HL-LHC~\cite{Astrand_2026, de2017fast, duarte2018fast, Govorkova:2021utb, schaefer2025advancing}. In gravitational-wave observatories, neural networks and reduced-order models (compressed representations of waveform templates that enable fast evaluation across parameter space~\cite{Cannon:2011vi,smith2020massively}) are investigated to accelerate matched filtering and low-latency detection across large template banks, supporting rapid multi-messenger alerts~\cite{george2018deep,raikman2024gwak}. 

\paragraph{Expensive simulations.} Generating synthetic data represent perhaps the most severe computational bottleneck across all domains. In collider physics, full detector simulation using GEANT4~\cite{agostinelli2003geant4} requires 1-10 seconds per event, and comprehensive analyses may need $10^9-10^{10}$ simulated events to match statistical power of data and explore systematic variations. Generative models—GANs, normalizing flows, diffusion models—can achieve speedups of 3-5 orders of magnitude by learning to sample directly from the distribution of simulated observables without explicitly tracking particles through detector materials~\cite{Krause:2024avx, anderlini2021machine, touranakou2022particle,javurkova2024fast,mazurek2025machine}. In cosmology, emulators trained on suites of expensive N-body and hydrodynamic simulations enable rapid prediction of observables as functions of cosmological parameters, making otherwise intractable parameter inference computationally feasible~\cite{doi:10.1073/pnas.1821458116,rodriguez2018fast,kamdar2016machine}. However, these fast surrogates must preserve critical correlations between observables and correctly propagate the effects of systematic uncertainties—requirements that demand rigorous validation before deployment in scientific analyses~\cite{Kansal:2022spb,10.1088/2632-2153/ae73e3}. This topic is discussed in details in the VERaiPHY review on Generative Models and Statistical Validation~\cite{diefenbacher2026generative}.


\paragraph{Intractable likelihoods.} The lack of close form likelihood models raise computational challenges at the stage of statistical inference, both to the practice of hypothesis testing and parameter estimation. Traditional methods like Markov chain Monte Carlo~\cite{metropolis1953equation,hastings1970monte} or nested sampling~\cite{skilling2006nested} require evaluating the likelihood at each step, and if each evaluation requires running expensive simulations, comprehensive parameter space exploration becomes prohibitively slow. \textit{Simulation-based inference (SBI)} methods~\cite{cranmer2020frontier}—neural likelihood estimation, neural posterior estimation, neural ratio estimation— amortize this cost by learning surrogate functions from a training set of simulations. In cosmology, field-level neural posterior estimation enables inference directly from high-dimensional maps of cosmic fields rather than compressed summary statistics, potentially extracting more information from observations. In collider physics, neural ratio estimation enables goodness of fit tests to detect new physics without specifying a signal model, and more traditional signal-specific parameter inference from a wider base of raw level observables. These methods make previously intractable inference problems solvable but require careful validation of their coverage and calibration properties to ensure they produce scientifically reliable uncertainty estimates.
This topic is further discussed in the VERaiPHY review on Simulation-based Inference with Machine Learning~\cite{max2026simulation}\\
\subsection{
New emerging AI paradigms}\label{sec:open}

Beyond accelerating existing workflows, AI is enabling qualitatively new approaches to scientific discovery. 
We move in the direction of \textit{self-driving science}, autonomous experiments and autonomous data analyses.

\paragraph{Fully differentiable workflows and optimal experimental design.} Traditional experimental designs are tailored to a specific signal hypothesis exploiting supervised methods of optimization. While proved successful in multiple situations (Higgs boson discovery~\cite{ATLAS:2012yve,CMS:2012qbp}, gravitational waves discovery~\cite{PhysRevLett.116.061102}), existing optimization schemes are far from optimal as the various components of the experimental workflow are tuned sequentially. In colliders, for instance, detector geometry is chosen based on physics requirements, then trigger algorithms are designed for that detector, data selections are optimized based on the stored raw information and statistical analyses are tuned on the available observables. In cosmology, survey strategies (which sky regions to observe, exposure times, filters) are determined by forecasting expected constraints. Recent work on differentiable programming enables end-to-end optimization~\cite{dorigo2023toward,aehle2025progress}. By making simulation, reconstruction, and inference differentiable with respect to design parameters, gradient-based methods can propagate sensitivity requirements backward. For collider detectors, one could optimize material composition, particles reconstruction, triggers, and even additional constraints such experimental costs of environmental impact by backpropagating through the full analysis to maximize discovery significance for a targeted signal hypothesis under budget constraints. For cosmological surveys, one could optimize observing strategies by propagating constraints on dark energy through the full inference pipeline. For gravitational wave detectors, parameters like mirror coatings or arm lengths could be optimized for specific science targets. While still at its early stages, this paradigm could transform how future experiments are designed, moving from sequential steps of optimization based on partial knowledge to an end-to-end optimization exploiting the complete information on the problem. One of the challenges of differentiable workflows is that many steps of the simulation process and data analysis are discrete, such as the number of layers and detector components and the clustering association of detector measurements to reconstructed particles. Approximations to emulate discrete choices as continuous decision boundaries are starting to emerge and careful studies to understand the validity of the approximations will be crucial to future autonomous systems.   

\paragraph{Active learning.} Most physics experiments operate in largely open-loop fashion: data collection strategies are determined before observations begin. Active learning flips this paradigm by adaptively directing resources toward regions of highest uncertainty or discovery potential. Applications span domains: in collider physics, adaptive trigger strategies could allocate bandwidth based on live anomaly scores~\cite{ATLAS:2022ttt}; in cosmology, robotic telescopes could prioritize follow-up observations of transients or peculiar galaxies identified by ML; in gravitational wave astronomy, detector configurations could adapt to observed source populations; in dark matter searches, scanning strategies could focus on parameter regions showing hints of excess. Challenges that affect several scientific areas are: (1) balancing exploration (searching new parameter space) and exploitation (refining promising signals); (2) ensuring interpretability of algorithmic choices, and (3) validating that adaptive strategies don't introduce selection biases that invalidate statistical inference.

\paragraph{Transfer Learning and Physics-Inspired Foundation Models.} Traditional machine learning workflows in fundamental science train task-specific models from scratch for each new analysis. Transfer learning changes this paradigm by developing general-purpose representations that can be efficiently adapted across experiments, observables, and even scientific domains. Unlike computer vision or natural language processing, where data naturally conform to regular formats such as images or text, scientific experiments record highly heterogeneous data: detector hits and reconstructed particles in colliders, galaxy catalogs in astronomical surveys, time-series measurements in gravitational wave detectors. This diversity has driven the development of physics-inspired architectures, incorporating domain-specific inductive biases—symmetries under rotations, permutations, or Lorentz transformations—to improve generalization.

Recent experimental campaigns have recorded unprecedented data volumes: billions of particle collisions at the LHC, millions of galaxy spectra from DESI~\cite{desi2026data}, hundreds of gravitational wave candidates from LIGO~\cite{abbott2023gwtc, abac2026gwtc}. These scale of the data has motivated foundation model approaches analogous to pre-training in language models~\cite{Feickert:2021ajf,Birk:2024knn,Hallin:2025ywf,Harris:2024sra,Golling:2024abg,Leigh:2024ked,Bardhan:2025icr}. Unlike text data, physics benefits from high-fidelity simulations encoding theoretical knowledge. Emerging pretraining strategies exploit such as supervised pretraining using simulation labels~\cite{Vigl:2024lat,Qu:2022mxj,Mikuni:2025tar,Mikuni:2024qsr} and contrastive learning~\cite{Lu:2024ict, Sheldon:2024sbe,Harris:2024sra, Metzger:2025ecl}. Early results demonstrate transfer across collision systems in particle physics and even between scientific domains~\cite{Mikuni:2025ocp,Elsharkawy:2026kwp}—suggesting data-rich experiments like the LHC could provide representations transferable to data-limited fields where simulation is computationally prohibitive.

However, creating foundation models for scientific data remains a challenge: (1) identifying optimal pretraining objectives that capture domain-specific physics while remaining flexible for downstream tasks; (2) quantifying when transfer is valid versus when domain shift invalidates learned representations; (3) ensuring interpretability of learned features to maintain scientific insight; and (4) developing theoretical frameworks to predict transferability across experimental conditions and physical regimes.

\noindent The challenges of building Foundation Models as well as more broadly meaningful data representations for fundamental physics purposes are further described in the VERaiPHY review on Representation Learning in Fundamental Physics~\cite{klein2026representation}.

\paragraph{Anomaly detection and signal-agnostic searches.} Signal-specific searches (such as supersymmetric particles in collider data, primordial non-Gaussianity in CMB maps, or specific WIMP masses in dark matter detectors) are powerful when theory provides clear guidance. However, due to their high specificity, narrow searches may completely  miss unexpected new phenomena. Anomaly detection frameworks that progressively lift theoretical constraints and favor learning structures directly from data can search for generic statistical deviations without committing to a specific physical model. In collider physics, these methods flag events inconsistent with the Standard Model predictions. In cosmology, they can identify unexpected structure in galaxy surveys or anomalous regions in CMB maps. For gravitational waves, anomaly detection helps identify new source populations not covered by existing waveform catalogs. 
Recent progress in representation learning has enabled the extraction of informative latent representations with unsupervised or semi/weakly-supervised techniques, like contrastive learning~\cite{Metzger:2025ecl}, that act as general purpose summary statistics (or foundation models) for multiple inference and testing tasks, widening the scope of model-agnostic searches to complex input data modalities (point clouds of particles, time series, highly structured images and graphs)~\cite{bright2026autoscidact}.
The open challenges shared across domains include (1) interpreting what makes observations anomalous and how this can inform the generation of new physical theories; (2) controlling false discovery rates in the presence of systematic uncertainties~\cite{dAgnolo:2021aun,tong2026covariant}; and (3) controlling false discovery rates when searching over many possible anomaly types simultaneously, where the look-elsewhere effect~\cite{davies1977,davies1987,Vitells:2011da} — governed by the data resolution and range as well as the (ML) model fitting the data — and the correlation structure of multiple tests on the same data must be jointly accounted for when calibrating statistical significance~\cite{grosso2025multiple,hein2025}.
The reader is referred to the VERaiPHY review ``Model-Agnostic Signal Discovery with Machine Learning: Bridging the Gap Between Theory and Practice"~\cite{amram2026model} for an extensive discussion of the topic.

\paragraph{Agentic AI}
More recently, Large Language Models (LLMs) are beginning to enter experimental workflows as interfaces to unstructured, text-based information. Their main utility lies in accessing, summarizing, and contextualizing knowledge encoded in documentation, logbooks, configuration files, software repositories, and prior literature, as well as in assisting with code development and debugging. In this role, LLMs support scientists by lowering the practical barriers to interacting with complex experimental infrastructures, rather than by replacing established numerical, statistical, or control algorithms.

Within experimental pipelines, LLMs can act as high-level assistants that help navigate analysis options, suggest relevant software tools or configuration changes, and translate human intent into executable code or workflow modifications. When coupled to tool-use interfaces—such as code execution environments, database queries, or job-submission systems—LLMs can help automate routine tasks, monitor logs and diagnostics, and flag anomalous behavior for human inspection. Importantly, decisions affecting data acquisition, inference, or discovery remain governed by explicit statistical procedures and human oversight. In this sense, LLMs function as productivity-enhancing mediators between humans and complex experimental systems, facilitating access to information and orchestration of workflows.\\

Beyond interfacing with documentation, LLMs are being developed as autonomous agents capable of end-to-end scientific analysis~\cite{lu2024ai}. 
Recent efforts have taken early steps toward LLMs orchestrating physics analysis workflows — interpreting physics questions, selecting datasets, generating and executing analysis code, and producing plots
~\cite{menzo2025heptapod,mcgreivy2025seeing,badea2026agentic, Moreno:2026mqk,diefenbacher2025agents}. These systems can navigate complex analysis frameworks (ROOT~\cite{ROOT_NIMA_1997}, Awkward Array~\cite{Pivarski_2020_Awkward_Arrays}, Coffea~\cite{CMS:2020kpn}), interface with experiment-specific data formats, and implement standard statistical procedures without human intervention at each step~\cite{getphysicsdone,getanalysisdone,autoresearch}. 
Nevertheless, fundamental challenges remain before these systems can meaningfully advance scientific discovery: ensuring physical correctness of generated code, maintaining statistical rigor throughout automated pipelines, and providing interpretable, verifiable reasoning at each step. Fully autonomous, end-to-end workflows that iterate reliably on intermediate results remain for now an open challenge.

The key capability is closed-loop reasoning: the agent examines outputs, diagnoses issues (missing branches, incorrect selections, statistical fluctuations), modifies code, and reruns until obtaining physically sensible results. This enables automated exploration of analysis variations—systematic uncertainty studies, background estimation strategies, selection optimization—that would be tedious for humans to enumerate manually. When combined with tool access to computing clusters, databases, and version control systems, such agents can execute analyses from natural language descriptions to publication-ready results.

However, this autonomy introduces critical verification challenges. Agents may produce statistically valid-looking results through incorrect physics reasoning, introduce subtle bugs in generated code, or optimize metrics that don't align with scientific goals. The irreversibility of certain analysis choices (unblinding, combinations of datasets) means errors can invalidate months of work. Similar to blind analyses~\cite{klein2005blind}, it is essential that agentic systems use only simulated samples when performing optimization routines, while reserving the observed data for the final statistical test. This separation prevents the optimization procedure from adapting to statistical fluctuations in the data, thereby reducing the risk of biased inference. In addition, applying the workflow to data only at the final stage provides an important validation step, helping to identify potential bugs or unintended artifacts introduced by the agentic workflow. Moreover, as analyses grow complex, verifying agent-generated workflows becomes as challenging as writing them manually—the verification bottleneck shifts rather than disappears. This stands in sharp contrast to domains such as AI for mathematics, where formal proof assistants like Lean~\cite{moura2021lean} provide an independent, mechanistic verification layer that has been a crucial ingredient in recent progress: a proof is either valid or it isn't, and this hard boundary allows agents to explore aggressively while failures remain detectable and recoverable. No analogous verification infrastructure exists for physics analyses, where correctness is probabilistic, domain-specific, and often only assessable in hindsight — making the development of such scaffolding one of the central open problems for trustworthy agentic science.

Current systems require substantial human oversight: reviewing generated code for correctness, validating intermediate distributions against physics expectations, and ensuring statistical procedures match intended inference goals. The promise is not full automation but human-AI collaboration where agents handle routine implementation while physicists focus on verification (interpretation, validation, and control)~\cite{miromind2026mirothinker}. Realizing this vision requires advances in both agent reliability and interpretability, ensuring autonomous systems produce not just correct numbers but understandable, reproducible science.

\section{When verification matters}\label{sec:whenwhy}
Section~\ref{sec:workflow} gave an overview of various ways machine learning is proving its power to enhance statistical analysis of fundamental physics data providing scalable and enhanced solutions to data selection, simulation and inference. 
The key question is not whether ML can accelerate or enhance each stage, but whether it can do so while preserving—or ideally improving—the statistical properties required for valid scientific inference. 
Under which conditions can we trust ML outputs to inform scientific conclusions? The answer depends critically on where and how ML is deployed within the statistical workflow introduced in Section~\ref{sec:workflow}.
\textit{Verification requires context.}

The VERaiPHY initiative reviews ML methods and their verification in a set of nine complementary contributions. 
This section connects to these topics and help contextualizing them within the statistical workflow for discovery, while discussing the principles for determining when and which rigorous verification methods are essential in various circumstances.

\subsection{When is it acceptable to be ``wrong''?}
Not all stages of the scientific workflow require the same level of fidelity from ML methods. Understanding when imperfect models are tolerable—and when they are not—is essential for allocating verification effort appropriately.

\paragraph{A wrong summarization is just suboptimal.} 

At the data summarization stage, ML methods transform raw detector signals into features or observables used in downstream analyses. A \textit{suboptimal choice of features}—one that discards relevant information or fails to compress optimally—will reduce statistical power (wider confidence intervals, lower discovery sensitivity) but will not introduce bias or compromise the validity of inference procedures, provided the transformation itself does not depend on unknown parameters that must be estimated from data.
For example, in collider physics, a jet tagging algorithm~\cite{ATLAS:2025dkv,CMS-DP-2024-066} that achieves 60\% signal efficiency instead of 80\% will require more data to reach the same statistical significance, but the resulting discovery claim remains valid if the selection efficiency is properly accounted for in the statistical model. In lattice calculations~\cite{Aarts:2026zzr}, AI models used to increase efficiency of phase space sampling may require more trials, leading to slower algorithms, but does not bias the calculations downstream. Similarly, in cosmology, a galaxy shape measurement algorithm that adds noise to shear estimates will degrade weak lensing constraints on dark matter, but does not bias parameter inference if the measurement uncertainties are correctly characterized. In these contexts, being ``wrong'' means being suboptimal rather than being systematically biased.

\paragraph{Generative models can be calibrated.}
ML-based fast simulation or generative models produce outputs that never perfectly match the true data distribution—they are ``wrong'' in the sense of being biased or misspecified relative to the ground truth. However, such models can still serve validly in the scientific workflow if their biases can be characterized and corrected through \textit{calibration procedures}.
The key distinction is between using a biased model directly for inference versus using it as a computational tool whose outputs are subsequently calibrated against data. For instance, a GAN-based calorimeter simulation might systematically mismodel the tails of energy distributions. If this model were used directly to estimate signal efficiencies or background rates, the bias would propagate to final results. However, if the generative model is used to produce large samples that are then corrected to match observed data distributions in control regions, the initial bias can be corrected. This is analogous to traditional simulation-based analyses where Monte Carlo generators with known deficiencies are calibrated using data-driven corrections.
The critical requirement is that calibration must be performed using independent data (not used in training) and must demonstrably reduce systematic differences between simulation and data to levels smaller than statistical uncertainties. Moreover, residual systematic uncertainties after calibration must be propagated through the analysis. Calibration transforms a ``wrong'' model into a useful tool, but does not eliminate the need for uncertainty quantification—it shifts the source of uncertainty from model fidelity to the calibration procedure itself.

\paragraph{Exploratory analysis can generate suggestions (not claims).} 
ML methods are used for exploratory data analysis, preliminary anomaly scans, or \textit{hypothesis generation} operating outside the formal inference workflow. Here, being ``wrong'' carries minimal risk because these applications do not directly produce publishable statistical claims. A clustering algorithm that identifies interesting substructure in data, even if it occasionally groups events spuriously, can still guide physicists toward regions worth investigating more carefully with rigorous methods. An ML-based anomaly score that flags unusual events serves its purpose if it draws attention to potentially interesting phenomena, even if some background processes are also identified as anomalous.
The crucial safeguard is that exploratory findings must be validated through independent, rigorous analyses before being claimed as discoveries. A ``wrong'' exploratory tool serves as a suboptimal hypothesis-generating engine, namely the chance of a real signal to be rigorously discovered are lower than the expected false positive rate. This is not an ideal and desirable scenario, but the downstream proper statistical test suffice to impede wrong claims. 
Critically, this testing must account for the look-elsewhere effect: if an ML method scans data to identify interesting regions or hypotheses, subsequent statistical tests of those hypotheses on the same data will suffer from post-selection bias~\cite{kuchibhotla2022postselection}. Properly, the hypothesis generated from exploratory analysis should be tested on an independent dataset not used in the exploration phase, or appropriate corrections for multiple implicit comparisons must be applied. This mirrors how physicists conduct blinded analyses, where signal regions are defined using simulation or control regions before examining actual data in those regions. Failing to account for the selection bias introduced by exploration can lead to spurious discoveries—the data will appear more significant than warranted because the hypothesis was constructed to fit fluctuations in that specific dataset.

\paragraph{Important caveats: when being wrong becomes unacceptable.} This tolerance for imperfection breaks down in several scenarios:\\

\textit{Hidden dependencies on systematics.} If a learned transformation's behavior depends on uncalibrated detector properties, beam conditions, or other systematic effects that vary in ways not captured by the statistical model, apparent harmless suboptimality becomes systematic bias. For instance, if the efficiency of a jet tagging algorithm in collider experiments depends on pileup (number of interactions per beam crossing) and this dependence differs between the training simulation and the test data, fixed efficiency corrections will be incorrect without the proper calibration step.\\

\textit{Exploitation of simulation artifacts.} If feature extraction or classification algorithms learn to exploit differences between simulation and data (rather than signal-vs-background differences), they will perform well in validation on simulated data but fail on real data in unpredictable ways. This is particularly insidious because standard validation procedures may not catch it.\\

\textit{Simulation imperfections in simulation-based inference.} In simulation-based inference, simulation samples under different hypothesis are used to extract physics parameters from measured data. However, if the simulations are not able to accurately capture all of the data components, the extracted parameters may be biased, or uncertainties may have the wrong coverage.\\

\textit{Uncharacterized calibration failures} Calibration procedures assume that biases are smooth and can be corrected through reweighting or functional fits. If a generative model has discontinuous support, mode collapse, or produces unphysical outputs in rare corners of phase space, calibration may not correct these issues, and they may go undetected until appearing in final results.\\

\textit{Feedback into training.} If ``wrong'' models are used to generate training data for other ML components, errors can compound. For instance, using a biased fast simulation to train a classifier, then using that classifier's outputs in an analysis, creates a chain where a potential generation bias is amplified rather than corrected.\\

The principle is: being suboptimal is acceptable as long as it induces the same efficiency in both the data and the reference samples; being biased is acceptable when biases can be calibrated out and residual uncertainties properly quantified; being wrong in ways that create systematic errors evading standard checks is never acceptable.

\subsection{When Must Uncertainties be Rigorously Quantified?}
\begin{figure}[h]
    \centering
    \includegraphics[width=0.9\linewidth]{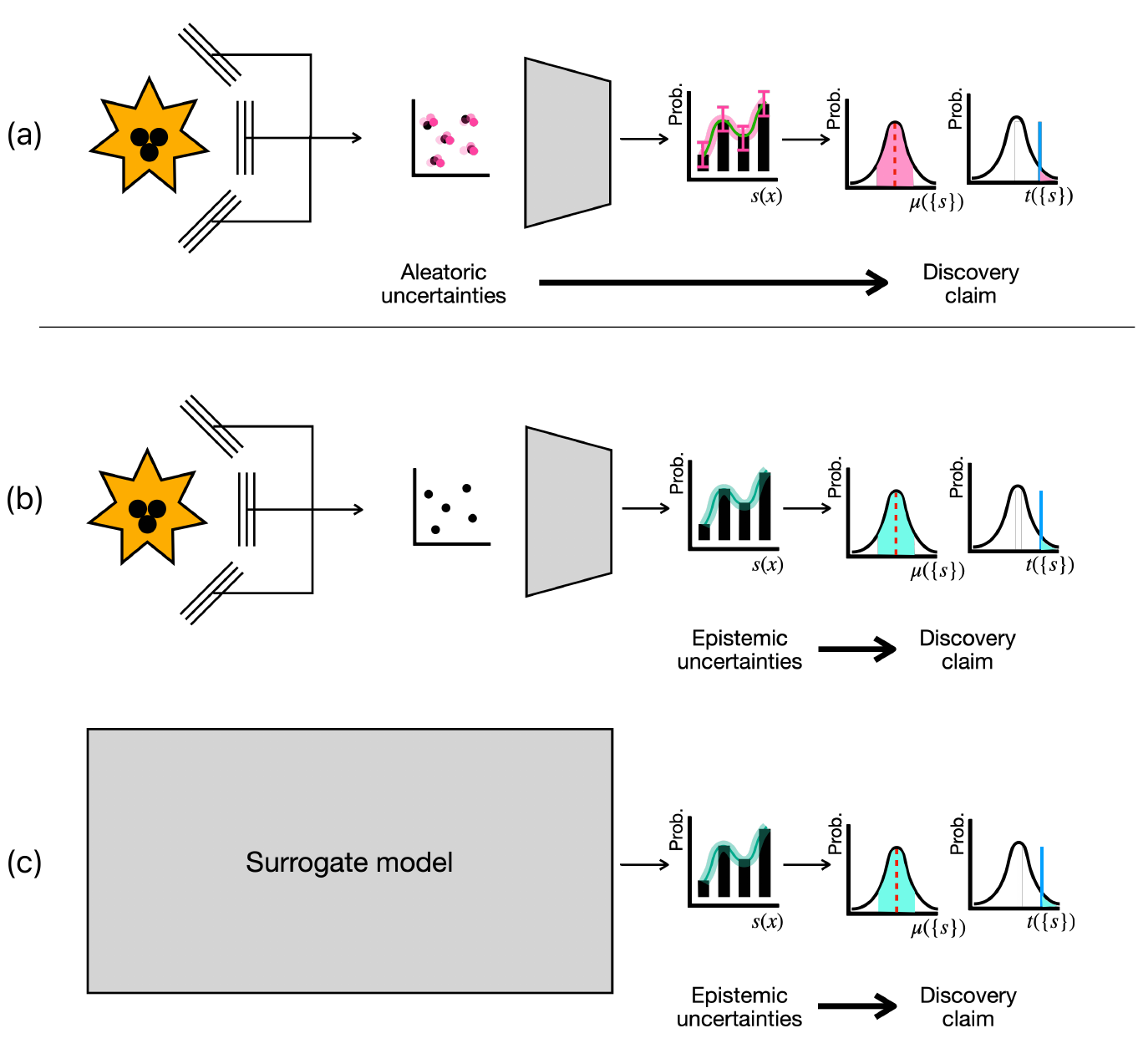}
    \caption{\textbf{When do we need to estimate uncertainties and which kind.} (a) Aleatoric uncertainties must be propagated from the collected data all the way to the discovery claim; (b) Epistemic uncertainties affecting the statistical model of the summary statistics must be included in the model as well; (c) the epistemic uncertainties affecting a surrogate model replacing a physical model for data generation must be propagated into the inference step that lead to scientific discovery.}
    \label{fig:when_uq}
\end{figure}

As discussed in detail in the VERaiPHY review on Uncertainty Quantification~\cite{haussmann2026uncertainty}, in the statistical workflow for scientific discovery there are uncertainties of different nature and origin. We primarily distinguish between \textit{aleatoric} and \textit{epistemic} sources of uncertainties; the former are uncertainties on the data generating process, including systematic and statistical ones due to finite sampling or experimental limits/imperfections; the latter are uncertainties on the statistical model describing the collected data in its final summary statistics.
Aleatoric uncertainties should always be carefully propagated from the input data representation to the final summary statistics and accounted for in the inference stage. Hence, any ML tools adopted in the statistical workflow should correctly propagate them as well (Figure~\ref{fig:when_uq} (a)).
The epistemic uncertainties of a ML model become relevant when the latter is used to build a robust statistical model of the data. Within the discovery workflow, this can happen in two stages: (1) when the summary statistics are modeled to perform inference (Figure~\ref{fig:when_uq} (b)); and (2) at the simulation level, if a ML generative model is used as a surrogate of a real physics model (Figure~\ref{fig:when_uq} (c)). 
Uncertainty propagation is non-negotiable for inference. 
If the ML model is not part of the statistical model used for inference/testing then it doesn't need epistemic uncertainties.




\subsection{When are Interpretability and Explainability Essential?}
As discussed in detail in the VERaiPHY reviews on Interpretability \& Explainability~\cite{Gambhir2026}, \textit{interpretability} refers to understanding how a model makes decisions and being able to anticipate its behavior, while \textit{explainability} refers to articulating why specific outputs were produced in domain-meaningful terms. Their necessity depends on where and how ML is deployed in the scientific workflow.

\paragraph{When Interpretability is Essential}
Interpretability becomes critical whenever ML systems operate in feedback loops or make decisions that directly influence the scientific process. This is essential for autonomous decision-making and experimental control, where scientists must verify systems optimize for scientific goals rather than exploiting artifacts; for monitoring and intervention, allowing detection of decisions that conflict with broader scientific or safety constraints; and for learning from AI, enabling the system to teach scientists about data structure rather than merely serving as an opaque tool.
Real-time trigger and selection systems represent another critical application. While individual event decisions need not be explainable, the overall trigger strategy must be interpretable enough that physicists can verify alignment with physics goals, assess coverage across phase space, and understand potential selection biases in subsequent analyses.
In practice, many physics analyses involve iterative refinement where domain experts examine intermediate results and guide subsequent choices. Interpretable ML enables this collaboration by allowing physicists to identify unexpected behavior, diagnose issues, and make informed decisions about trusting or revising ML components.
\paragraph{When Explainability is Essential}
Explainability becomes critical when ML outputs must be translated into physical understanding or drive scientific decisions requiring justification.\\

\textit{Anomaly detection and discovery} is a good example. If an ML-based anomaly detector flags a statistically significant excess, explainability determines whether this could lead to a discovery or remains a statistical curiosity. The detector must explain what makes flagged events anomalous: their distribution in phase space, unusual correlations between observables, or consistency across detector subsystems versus correlation with instrumental effects. Without explainability, a significant deviation (small p-value) is insufficient—physicists need to understand the anomaly's nature to determine whether it warrants follow-up, suggests new physics or detector mismodeling, and how to design refined searches. An explainable detector that reveals excess events concentrated in specific kinematic regions with particular correlations provides actionable information; an opaque detector returning only a p-value does not.\\

\textit{Model validation and diagnostics} similarly require explainability. ML-based goodness-of-fit tests must explain what aspects of data are poorly described: specific observables, kinematic regions, or variable correlations. This guides whether to refine theoretical models, improve calibration, or investigate systematics.\\

\textit{Guiding theoretical development} represents an opportunity where explainability transforms ML from computational tool to source of scientific insight. A learned classifier that outperforms physics-motivated variables, if explainable, can reveal previously unidentified correlations in data structure, suggesting new observables or inspiring theoretical work on their physical origin.

\paragraph{When Interpretability/Explainability are Optional}
It is important to recognize that not all ML applications require interpretability or explainability:\\

\textit{Event-level classification in validated workflows.} Once a tagging algorithm has been rigorously validated (calibrated on data, tested for systematic dependencies, verified to maintain efficiency across phase space), its event-by-event decisions need not be interpretable for use in downstream analysis. The analysis only requires knowing the overall signal efficiency and background rejection, properly accounting for systematic uncertainties in these quantities. Individual tag decisions can remain opaque.\\

\textit{Computational acceleration with verified outputs.} Fast simulation, when used purely to accelerate computation, does not require explainability at the level of individual generated events. What matters is that the ensemble of generated events correctly reproduces the statistical properties of full simulation: correct distributions, correlations, and response to systematic variations. If these properties are verified through rigorous validation tests, the internal workings of the generative model can remain opaque.\\

\textit{Exploratory data analysis.} ML used for preliminary data exploration, visualization, or hypothesis generation operates outside the formal inference workflow. While interpretability may enhance these applications by providing scientific insight, it is not required for their functional purpose.
\section{Fundamental limitations to AI-driven science}\label{sec:limits}
The preceding sections have established when and why verification matters (Section~\ref{sec:whenwhy}) and identified new research direction in AI for science (Section~\ref{sec:open}). However, certain fundamental limitations constrain what can be achieved through machine learning, regardless of algorithmic advances, computational resources, or verification rigor. Recognizing these intrinsic boundaries is essential for maintaining realistic expectations about ML capabilities and for designing experiments and analyses that work within rather than against these constraints.

\subsection{Inductive Bias is Unavoidable}
\paragraph{No learning without assumptions.} The No Free Lunch theorems~\cite{wolpert1995no,wolpert1996lack,wolpert2002no} formalize the fundamental principle that there is no universal learner: no learning algorithm can be good  at all problems. 
Effective learning requires inductive bias, built-in assumptions about what patterns are likely to be meaningful for a specific task or data structure. 

\paragraph{Sources of inductive bias.}
Inductive bias manifests at multiple levels of model design. Neural network architectures embody structural assumptions: convolutional networks assume local spatial correlations; graph neural networks assume relationships defined by graph connectivity; transformers assume patterns accessible through attention mechanisms. Physics-informed architectures explicitly encode symmetries—Lorentz invariance in particle physics, rotational symmetry in cosmology, gauge invariance in field theory. Beyond architecture, bias appears in loss function design (which features to prioritize), activation functions (smoothness, boundedness, nonlinearities), regularization penalties (L1 favoring sparsity, L2 favoring small weights), data augmentation strategies (which transformations preserve meaning), and optimization choices (learning rates, batch sizes, gradient clipping).
Bias can also be induced at a higher level in the data content and structure by designing the data preprocessing (data filtering, compression, parametrization).

\paragraph{The apparent universality of modern architectures.}
The impressive success of transformers across many domains (text, images, audio, physical data), and of convolutional and Recurrent NNs before then, may seem to contradict the No Free Lunch theorem, suggesting they have discovered a bias-free universal learner. 
Recent theoretical work shed light into this apparent paradox. Deep neural networks generically exhibit a strong simplicity bias: the parameter-function map is exponentially biased toward functions of low Kolmogorov complexity~\cite{valle2018deep}, a property shared across architectures (CNNs, ResNets, transformers) and arising from both architectural components (normalization, residual connections, activation functions) and optimization dynamics (SGD favoring flatter minima)~\cite{huh2021low, goldblum2023no}. Critically, natural data across modalities—text, images, physical measurements—tends toward low Kolmogorov complexity: it is structured, compressible, and far from random~\cite{finzi2026entropy}. Deep networks cross-domain success reflects this alignment: their soft biases toward low-complexity solutions match properties shared across natural datasets~\cite{goldblum2023no}. Even ``universal" architectures succeed precisely because their implicit biases align with the structure of problems we care about—inductive bias remains unavoidable.

\paragraph{Choosing bias, not eliminating it.} The question is never whether to include inductive bias, but which bias to choose and how explicitly to encode it. For example, rotational symmetry in images can be enforced architecturally through equivariant networks~\cite{cohen2016group}, which guarantee the constraint holds for any input, or expressed softly through data augmentation~\cite{dangovski2021equivariant}, which encourages the model to learn the symmetry from examples but does not guarantee it. Architectural encoding is stricter and more sample-efficient but less flexible; augmentation-based encoding is easier to implement and more adaptable but may fail to generalize perfectly. 
More generally, choosing how to encode knowledge in the learning process involves fundamental trade-offs:\\

\textit{Soft vs. hard constraints:} Expected symmetries can often be broken in experimental data by finite detector resolution and measurement capabilities, placing desired functions outside the model's constrained space. 
Even when symmetries are expected to hold, 
strict enforcement can be detrimental~\cite{nabat2025learning,langer2024probing}.  When searching for deviations from known theories, ML methods that perfectly enforce symmetries cannot discover new physics violating those symmetries.  Moreover, from a technical standpoint, constrained optimization may slow convergence or prevent finding optimal solutions. Soft constraints allow for a larger space of solutions, allowing symmetry breaking during the optimization stage as well as in the final solution.\\

\textit{Domain knowledge vs. data-driven learning:} Incorporating strong physics priors (conservation laws, known interactions) improves sample efficiency and interpretability but risks encoding incorrect or incomplete physics. Minimal-assumption approaches require more data and may learn spurious correlations but cannot be misled by incorrect priors.\\

\noindent This topic is treated in depth in the VERaiPHY review on Symmetry-Informed Machine Learning~\cite{spinner2026symmetry}.

\subsection{Observational and Experimental Constraints}

\paragraph{We can only learn from what we measure.} Experiments provide access to specific observables through particular measurement processes. Fundamental physics in unmeasured or unmeasurable regimes remains empirically inaccessible, regardless of ML sophistication.
This constraint manifests differently across domains:\\

\textit{Detector acceptance, coverage and resolution:} Detectors have gaps in angular coverage, energy thresholds, and signal identification capabilities. Physics processes producing signals outside detector acceptance cannot be directly reconstructed. ML can improve reconstruction within covered regions but cannot guarantee the generalization to uncovered phase space.
More fundamentally, this points to an intrinsic ceiling on ML performance in experimental physics that has no direct analog in many other ML application domains: the experimental apparatus imposes a hard information-theoretic limit on what can be learned from the data. Unlike natural language or vision tasks where scaling laws~\cite{kaplan2020scaling,hoffmann2022training} suggest continued improvement with more data and compute, physics analyses are ultimately bounded by the resolution, acceptance, and luminosity of the detector. Larger models or more sophisticated architectures cannot recover information that was never recorded — a systematic uncertainty floor that is irreducible regardless of algorithmic advances. This has direct implications for how scaling laws should be interpreted in physics contexts: gains that appear robust on benchmark tasks may plateau or saturate when confronted with the fundamental constraints of the measurement process.\\

\textit{Cosmic variance and sample size of one:} Cosmology observes a single realization of the Universe. Statistical inference must contend with cosmic variance—fundamental uncertainty from not observing all possible realizations of cosmic initial conditions. Forthcoming surveys will image billions of galaxies, but these all come from one Universe, limiting our ability to test certain cosmological hypotheses.\\

\textit{Luminosity and exposure time limits:} Rare process searches face irreducible limits from finite observation time. Luminosity (in colliders experiments) or exposure time (in astroparticle experiments) determine the statistical power of an experiment, and the overall evidence in favor or against a hypothesized statistical model.\\

\textit{Systematic uncertainties:} Beyond statistical limitations, systematic uncertainties--detector calibration precision, theoretical prediction accuracy, background modeling fidelity --establish floors below which improvements in statistical precision provide no additional physics reach. An analysis limited by 5\% systematic uncertainty based on detector resolution gains nothing from reducing statistical uncertainty to 0.1\% through better ML methods.\\

\textit{The degeneracy problem:} Some physics parameters or model components may be fundamentally unidentifiable from available observables~\cite{efstathiou1999cosmic}: different parameter combinations or model structures can produce identical observable distributions. ML cannot resolve such degeneracies without additional measurements or strong prior assumptions. For instance, in cosmology, certain combinations of cosmological parameters (geometric degeneracies) produce nearly identical observational signatures, requiring multiple independent probes and measurement strategies to break degeneracies.

\paragraph{Data Augmentation Cannot Create New Evidence.}
Data augmentation—creating additional training examples through transformations of existing data—is a ubiquitous technique in ML. However, augmentation cannot create genuinely new information; it can only make explicit what was implicit in the original data plus the assumptions encoded in the augmentation procedure. This is not merely a practical limitation but a consequence of the Data Processing Inequality~\cite{cover1999elements}: any deterministic or stochastic transformation of data cannot increase the mutual information between the processed output and the quantity of interest. Augmentation can reorganize and surface implicit structure, but the information-theoretic ceiling is set by the original data and the assumptions baked into the procedure itself. Despite not creating new evidence, data augmentation is a powerful form of inductive bias for machine learning tasks, explicitly exposing models to relevant piece of information that might be available only implicitly and thus hard to learn for a machine. 
In the following we list three relevant classes of augmentations:\\

\textit{Symmetry-based augmentation:} augmentations highlighting physical constraints. Rotating or reflecting images of galaxies, boosting particle collision events, or time-reversing trajectories (for time-reversal-invariant processes) applies known symmetries. This augmentation is valid because physics guarantees these transformations preserve physical content. However, it cannot create information beyond what symmetry already implied—it trades computational cost for data volume.\\

\textit{Simulation-based augmentation:} augmentations that exploit existing models of the data to produce additional training points.
When using traditional physics driven simulators to create additional training examples, one can produce augmentations according to different theory and nuisance parameters, making domain shifts relevant to the physics case explicit and learnable~\cite{PhysRevD.111.032010,5n77-ynsp}. It should be noted that when using generative AI models as a proxy for the true statistical model, assumptions about the data distribution are implicitly encoded in the generative model itself. If the model is wrong, augmentation amplifies model misspecification.
\\

\textit{Out-of-distribution extrapolation:}  augmentations that extrapolate beyond the support of the observed data (e.g., generating examples with parameter values not represented in training data) by making predictions, not creating new evidence. Such extrapolation can guide exploration but cannot substitute for actual measurements.\\

\subsection{Computational Bounds}

\textbf{Effective information is observer-dependent.}
Classical information-theoretic notions—such as Shannon entropy, Kolmogorov complexity, and Fisher information—are defined for idealized observers with unbounded computational power, unlimited memory, and access to arbitrarily complex operations~\cite{finzi2026entropy,xu2020theory}. Under these assumptions, information is an intrinsic property of data-generating processes, independent of the observer. In contrast, all practical machine-learning systems operate under strict computational constraints: finite precision arithmetic, limited memory, bounded training time, and restricted algorithmic complexity. As a result, what constitutes \emph{accessible} or \emph{usable} information becomes observer-dependent.

This recognition has motivated several complementary frameworks for reasoning about information under computational bounds. Ref.~\cite{finzi2026entropy} introduces the notion of \emph{epiplexity}, which quantifies structural information accessible to resource-limited learners, decomposing total information into learnable structure and residual \emph{time-bounded entropy} that remains indistinguishable from noise within a given computational budget. Ref.~\cite{xu2020theory} develops \emph{predictive V-information}, a variational extension of Shannon's information theory that explicitly accounts for modeling power and computational constraints, allowing information to be "created" through computation in violation of the classical data processing inequality. Resource-bounded variants of Kolmogorov complexity~\cite{longpre1986resource,allender2006power} provide computable approximations by restricting computation to polynomial time, enabling practical applications while preserving key theoretical properties. These frameworks share a common insight: for computationally bounded observers, the distinction between what is theoretically present in data and what can be practically extracted becomes fundamental.

Two implications are particularly relevant for physics applications of ML. First, computationally bounded learners are sensitive to the sequence in which data are presented: ordering effects can qualitatively change what can be learned within fixed computational budgets~\cite{finzi2026entropy}, with concrete consequences for curriculum learning, adaptive simulation strategies, and sequential experimental design. Second, while deterministic transformations cannot increase Shannon entropy or Kolmogorov complexity in an absolute sense, they can make previously inaccessible structure explicit for bounded observers~\cite{xu2020theory}. Running a simulation—such as a Monte Carlo event generator—transforms implicit structure into a computationally accessible form; similarly, learned surrogate models that trade exactness for speed may discard only fine-grained details inaccessible to downstream inference, preserving the epiplexity relevant for learning good data representations. This perspective reframes computational constraints as a fundamental limitation on scientific inference with ML, rather than a purely technical inconvenience.

\subsection{Verification Itself Has Limits}

Verifying that an ML method satisfies desired statistical properties (coverage, calibration, unbiasedness) requires ground truth or validation procedures that themselves rely on assumptions.\\ 

\textit{The missing ground-truth problem.}
We typically validate either on simulated data or on real data through data-driven techniques (using control regions, cross-validation, calibration on auxiliary measurements). The benefit of using simulations is that the ground truth is known by construction; on the other hand,  simulations often do not adequately represent real data. Similarly, control regions can fail to represent nominal data: the definition of control regions relies itself on assumptions on the properties of the data belonging to this region (for instance, absence of signal or absence of systematic shifts). In this context, calibration strategies are often derived as a function of fewer observables than the ones used to train algorithms, and may not be able to cover all the existing mis-modeling. The fundamental difficulty is that we cannot fully validate methods on scenarios we haven't yet observed, the very scenarios where new discoveries lie.\\

\textit{Adversarial robustness limits.} Adversarial examples demonstrate that ML models can fail catastrophically on inputs slightly perturbed from training distributions~\cite{goodfellow2014explaining}. While physically meaningless adversarial attacks (pixel-level perturbations designed to fool classifiers) are less concerning in physics than in vision applications, the underlying phenomenon—ML models learning shortcuts rather than robust features—remains relevant. Verification can identify many such failures, but providing formal guarantees of robustness over all possible distribution shifts may be intractable. Conversely, one may be tempted to train AI models to be explicitly robust against adversarial attacks. Such strategy could promote robustness to well known failure modes, but be powerless against unexpected ones (there is no hope to enumerate all the ways to circumvent controlled behaviors, especially in highly parametrized models). Moreover, such fine-tuning can significantly lower the performance of the AI model in favor of robustness. Additionally, one may want to train a model to be robust against partially known systematic effects, such as the differences between the specific choice of physics generator. While this choice ensures the model response is less sensitive to simulation choices, it does not necessarily make the results more robust, since the true  phenomena in nature may lie somewhere else that is not captured by either of the physics models. \\

\textit{Statistical guarantees under model misspecification.} Many verification procedures assume a correctly specified statistical model. However, all models are approximations. Assessing whether a method maintains desired properties under realistic model misspecification—where we don't know precisely how the model is wrong—presents fundamental challenges. Sensitivity analysis and stress testing can probe robustness to anticipated misspecifications, but unknown unknowns remain.\\

\noindent Approaches to tackle these challenges are discussed in the VERaiPHY review on Unknown Unknowns in Machine Learning for Physics~\cite{cruz2026unknown}.
\section{The Physicist of the Future}
\begin{figure}[t]
    \centering
    \includegraphics[width=1\linewidth]{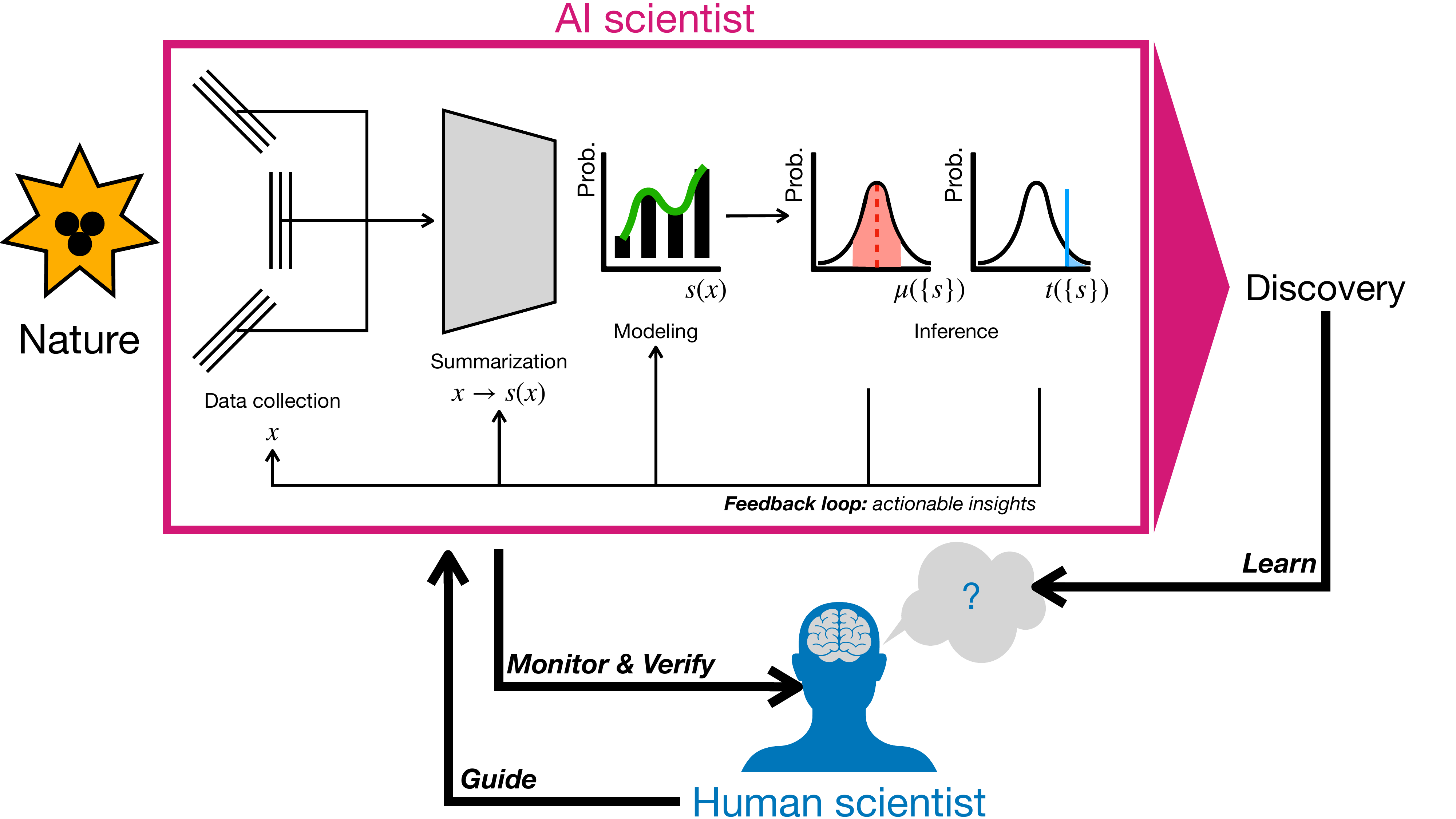}
    \caption{\textbf{The physicist of the Future.} We lean toward orchestration, monitor and verification of increasingly automatic pipelines.}
    \label{fig:future}
\end{figure}

The integration of semi-autonomous AI systems into the scientific discovery pipeline has the potential to radically \textit{accelerate} statistical inference, enabling physicists to probe fundamental phenomena at a scale and level of complexity previously unattainable. As increasingly sophisticated analysis tools become available at the push of a button, the role of the scientist will inevitably shift. Rather than spending most of their effort on implementing and optimizing statistical procedures, physicists and statisticians will devote more attention to formulating the right scientific questions, defining meaningful hypotheses, and determining which results should be combined and how.
In this setting, the central challenges will no longer be purely computational, but epistemic: questions of knowledge and understanding. How should one prioritize among a rapidly proliferating set of hypotheses and tests? When is automated inference trustworthy enough to guide experimental decisions or support discovery claims? How should uncertainty, bias, and model assumptions be monitored as AI systems increasingly operate in closed feedback loops with experiments? Interpretability, verification, and continuous validation will become core scientific activities, with human oversight focusing on understanding, controlling, and contextualizing the outputs of powerful but potentially imperfect learning systems (see Figure~\ref{fig:future}).

\paragraph{Studying AI as a physical system.}
Beyond deploying AI to solve physics problems, physicists are uniquely positioned to study AI systems themselves as complex dynamical systems exhibiting emergent phenomena~\cite{simon2026there}. Large neural networks trained via stochastic gradient descent can be viewed as high-dimensional, nonlinear dynamical systems evolving under optimization dynamics, with phase transitions in learning behavior, spontaneous symmetry breaking in learned representations, and collective phenomena arising from interactions among billions of parameters. The tools of statistical physics—mean field theory, replica methods, random matrix theory—have already provided insights into neural network training dynamics, generalization, and the structure of loss landscapes~\cite{bahri2020statistical}. Concepts from critical phenomena help explain scaling laws and emergence of capabilities at specific model sizes~\cite{hoffmann2022training,kaplan2020scaling,bahri_pnas}. Information-theoretic approaches characterize the compression and representation of data during learning. As AI systems grow in complexity and capability, understanding their behavior may require the same theoretical frameworks physicists have developed to study complex many-body systems, non-equilibrium statistical mechanics, and emergent collective behavior. This represents a natural extension of physics beyond traditional domains: just as physicists study phase transitions in magnets or information flow in quantum systems, the collective behavior of artificial neural systems presents rich phenomena amenable to physical modeling and theoretical understanding. Moreover, insights from studying AI as physical systems can feed back into better verification frameworks—if we understand the dynamical origins of phenomena like scaling laws, phase transitions in training, or mode collapse in generative models, we can develop more principled approaches to predicting and controlling AI behavior, complementing empirical validation with theoretical understanding.
An in-depth discussion of this topic can be found in the VERaiPHY review on Statistical Properties of Training \& Generation~\cite{lavie2026statistical}.

\paragraph{Teaching AI through verification.}
An emerging paradigm positions human verification as an integral training component rather than merely post-hoc validation. Reinforcement learning from human feedback (RLHF)~\cite{christiano2017deep,ouyang2022training} learns reward models from expert evaluations —transforming physicists' judgments about analysis quality, systematic uncertainty treatments, or scientific plausibility into optimization objectives that guide model improvement. Process supervision~\cite{lightman2023let,uesato2022solving} rewards valid intermediate reasoning steps rather than just final answers, naturally aligning with scientific methodology's emphasis on sound methodology. Constitutional AI~\cite{bai2022constitutional} embeds verification principles as training constraints, enabling self-correction against rules like conservation laws or dimensional consistency. Physicists' expertise in evaluating scientific soundness—distinguishing valid from flawed or incomplete reasoning, identifying violations of physical principles, assessing methodological rigor—becomes a powerful teaching mechanism that can encode scientific goals directly into AI systems through structured feedback~\cite{Thais2024}. Frameworks like VERaiPHY provide useful guidelines for which aspects of AI outputs require verification and what standards they must meet, which can inform both what physicists evaluate during training and how those evaluations are structured. However, this integration introduces new challenges: reward models can be gamed, human evaluators may be inconsistent, and as AI capabilities grow, ensuring the quality and reliability of verification judgments becomes critical—particularly when models potentially exceed individual expert capability, requiring collective oversight mechanisms.

\paragraph{The enduring value of learning analysis skills.}
Facing rapid growth of advanced reasoning tools based on AI, students and researchers may wonder whether to invest effort in learning skills that machines can perform faster. However, as emphasized throughout this review, developing and deploying sophisticated AI-based analysis methods requires meticulous verification and constant monitoring, tasks that demand human experts. We can teach machines how to properly perform data analysis only if we know how to do it ourselves. Education, development of critical thinking, and the ability to make informed decisions are skills we must cultivate and transfer to young generations of researchers to ensure safe, robust, and controlled use of new technologies.
In the past, the purpose of education has often being aligned with performing tasks efficiently and impeccably for examination—in other words, training students to resemble perfect machines in execution. Today, it should be clear that education's purpose is not to compete with machines but to nurture our distinctly human capabilities: curiosity, creativity, critical judgment, and the ability to formulate meaningful questions. Educational exercises are given not to see who solves them fastest, but to impart the scientific process and how to approach complex problems.
In this changing paradigm, future researchers will need deep understanding of entire analysis workflows to verify AI-generated results, identify issues with generated code, and propose improvements. While the specific methods employed to answer scientific questions may evolve as AI capabilities grow, the foundational knowledge necessary to understand the physics questions themselves, the structure of data analysis workflows, the statistical tests employed and their applicability remains essential. This understanding enables researchers not merely to execute analyses but to conceive future questions, evaluate whether automated procedures are trustworthy, and steer scientific progress with informed judgment. The goal is not efficiency alone, but the combination of computational power with human insight that ensures robust, verifiable, and meaningful discoveries~\cite{ThaisManuscript-THAAFS-2}.

\section{Concluding remarks}\label{sec:conclusions}

ML has become integral to the statistical workflows of fundamental physics, accelerating computation at every stage—from data acquisition and reconstruction through simulation, inference, and hypothesis testing—while enabling analyses at scales previously infeasible. As these systems grow increasingly autonomous, from adaptive triggers to agentic experimental control, the need for rigorous verification becomes not merely desirable but essential. This review has established basic guidelines to verification. First of, effective verification requires context: the requirements depend critically on where ML enters the workflow and what role its outputs play in scientific conclusions. Deterministic transformations in summarization can be suboptimal without invalidating downstream inference, while biased models in hypothesis testing directly corrupt discovery claims. Understanding this distinction—between what degrades performance and what compromises statistical validity—is fundamental to deploying ML responsibly. AI-driven discovery is not possible otherwise. However, even perfectly verified ML systems face irreducible constraints that can limit the discovery power of AI tools: finite computational resources bound what information can be extracted, limited data constrain what can be learned, and unavoidable assumptions could bias the learning strategy. Within these boundaries, physicists play an irreplaceable role: not merely as users of AI tools, but as designers of experiments that determine what can be discovered, as evaluators whose verification judgments guide training, and as interpreters who translate statistical findings into physical understanding. The responsible integration of ML into fundamental physics requires recognizing both its transformative potential and its fundamental limitations, ensuring that AI-driven discoveries rest on sound statistical and epistemic foundations.

\paragraph{General guidelines for AI verification:}
In the following we summarize the key take-aways of the review. We refer the reader to the full set of VERaiPHY articles for an in depth series of guidelines for specific use cases.\\

\textit{What statistical question does my AI model answer?}
Before introducing a new ML method in your statistical analysis ask yourself two fundamental questions: (1) which stage of the statistical workflow does the model belong to? and (2) what statistical question does the model aim to address? Answering these questions would help you contextualize your research and guide you in the verification stages. \\

\textit{Design experiments for learning.} Recognize that ML cannot discover unobserved phenomena alone. Experimental design, detector optimization, and resource allocation determine what can ultimately be learned. ML should inform experimental design, and experiments should be designed with ML capabilities in mind, but any prediction outside of the training scope needs verification before discovery claims.\\

\textit{Embrace, don't fight, inductive bias.} Since bias is unavoidable, explicitly choose and justify it based on domain knowledge. For example, physics-informed architectures that encode well-established principles while maintaining flexibility where physics is uncertain represent principled approaches. When possible, controlling the mathematical properties of your model will help verify its behavior. Even scalability promoted by transformers is an inductive bias and may be a well justified choice given the amount of data and compute available.\\


\textit{Verify rigorously, but recognize verification limits.} Apply the verification frameworks discussed throughout the VERaiPHY series of reviews while acknowledging that verification cannot provide absolute guarantees. Multiple complementary verification approaches, stress testing under diverse scenarios, and community-wide scrutiny of high-stakes results provide defense in depth. In particular, always raise concern when a tool is deployed out-of-domain (e.g. on data that strongly differ from the ones used at training and during verification stages).\\

\textit{Maintain epistemic humility.} Claims based on ML should acknowledge both statistical uncertainties and the assumptions embedded in methods. Discovery claims require reproducibility, independent validation, and demonstration of robustness to methodological choices, standards that become even more important when ML mediates between data and conclusions.

\section*{Acknowledgements}
The authors thank Michael Kagan and Phil Harris for the inspiring discussions and Benjamin Nachman and Nick Wardle for insightful comments on the document. The authors would like to extend a particular acknowledgement to Ramon Winterhalder for coordinating the VERaiPHY initiative, and to Lydia Brenner, Louis Lyons and Tilman Plehn for initiating this effort in the first place.

\paragraph{Author contributions}
GG and VM are equal contributors and main authors to all aspects of this review article. LH provided advice about article scope and high-level content, and valuable feedback during the drafting and finalization of the manuscript.

\paragraph{Funding information}
GG acknowledges the financial support of the National Science Foundation under Cooperative Agreement PHY-2019786 (The NSF AI Institute for Artificial Intelligence and Fundamental Interactions, http://iaifi.org/), VM is supported by JST EXPERT-J, Japan Grant Number JPMJEX2509. LH is supported by the Excellence Cluster ORIGINS, which is funded by the Deutsche Forschungsgemeinschaft (DFG, German Research Foundation) under Germany’s Excellence Strategy - EXC-2094-390783311 and by the Project SciFM 05D25WO2  funded by the German Federal Ministry of Research, Technology, and Space (BMFTR).






\bibliography{references.bib}

@misc{cho2025neurips,
  author = {Cho, Kyunghyun},
  title = {From Benchmarks to Problems - A Perspective on Problem Finding in AI},
  howpublished = {Invited Talk, Conference on Neural Information Processing Systems (NeurIPS)},
  year = {2025},
  month = {December},
  address = {San Diego, CA},
  url = {https://neurips.cc/virtual/2025/loc/san-diego/invited-talk/109605}
}

@article{CMS:2020kpn,
    author = "Smith, Nicholas and others",
    editor = "Doglioni, C. and Kim, D. and Stewart, G. A. and Silvestris, L. and Jackson, P. and Kamleh, W.",
    collaboration = "CMS",
    title = "{Coffea: Columnar Object Framework For Effective Analysis}",
    eprint = "2008.12712",
    archivePrefix = "arXiv",
    primaryClass = "cs.DC",
    reportNumber = "FERMILAB-CONF-20-494-CMS-SCD, CMS-CR-2020-069",
    doi = "10.1051/epjconf/202024506012",
    journal = "EPJ Web Conf.",
    volume = "245",
    pages = "06012",
    year = "2020"
}

@inproceedings{Pivarski_2020_Awkward_Arrays,
  title = {Awkward Arrays in Python, C++, and Numba},
  author = {Pivarski, Jim and Elmer, Peter and Lange, David},
  booktitle = {Proceedings of the 24th International Conference on Computing in High Energy and Nuclear Physics (CHEP 2019)},
  journal = {EPJ Web of Conferences},
  volume = {245},
  pages = {05023},
  year = {2020},
  doi = {10.1051/epjconf/202024505023},
  url = {https://doi.org}
}

@article{ROOT_NIMA_1997,
    author = "Brun, R. and Rademakers, F.",
    editor = "Werlen, M. and Perret-Gallix, D.",
    title = "{ROOT: An object oriented data analysis framework}",
    doi = "10.1016/S0168-9002(97)00048-X",
    journal = "Nucl. Instrum. Meth. A",
    volume = "389",
    pages = "81--86",
    year = "1997"
}

@article{ATLAS:2022ttt,
    collaboration = "ATLAS",
    title = "{Active Learning reinterpretation of an ATLAS Dark Matter search constraining a model of a dark Higgs boson decaying to two b-quarks}",
    reportNumber = "ATL-PHYS-PUB-2022-045",
    year = "2022"
}

@article{PhysRevLett.116.061102,
  title = {Observation of Gravitational Waves from a Binary Black Hole Merger},
  author = {Abbott, B. P. and Abbott, R. and Abbott, T. D. and Abernathy, M. R. and Acernese, F. and Ackley, K. and Adams, C. and Adams, T. and Addesso, P. and Adhikari, R. X. and Adya, V. B. and Affeldt, C. and Agathos, M. and Agatsuma, K. and Aggarwal, N. and Aguiar, O. D. and Aiello, L. and Ain, A. and Ajith, P. and Allen, B. and Allocca, A. and Altin, P. A. and Anderson, S. B. and Anderson, W. G. and Arai, K. and Arain, M. A. and Araya, M. C. and Arceneaux, C. C. and Areeda, J. S. and Arnaud, N. and Arun, K. G. and Ascenzi, S. and Ashton, G. and Ast, M. and Aston, S. M. and Astone, P. and Aufmuth, P. and Aulbert, C. and Babak, S. and Bacon, P. and Bader, M. K. M. and Baker, P. T. and Baldaccini, F. and Ballardin, G. and Ballmer, S. W. and Barayoga, J. C. and Barclay, S. E. and Barish, B. C. and Barker, D. and Barone, F. and Barr, B. and Barsotti, L. and Barsuglia, M. and Barta, D. and Bartlett, J. and Barton, M. A. and Bartos, I. and Bassiri, R. and Basti, A. and Batch, J. C. and Baune, C. and Bavigadda, V. and Bazzan, M. and Behnke, B. and Bejger, M. and Belczynski, C. and Bell, A. S. and Bell, C. J. and Berger, B. K. and Bergman, J. and Bergmann, G. and Berry, C. P. L. and Bersanetti, D. and Bertolini, A. and Betzwieser, J. and Bhagwat, S. and Bhandare, R. and Bilenko, I. A. and Billingsley, G. and Birch, J. and Birney, I. A. and Birnholtz, O. and Biscans, S. and Bisht, A. and Bitossi, M. and Biwer, C. and Bizouard, M. A. and Blackburn, J. K. and Blair, C. D. and Blair, D. G. and Blair, R. M. and Bloemen, S. and Bock, O. and Bodiya, T. P. and Boer, M. and Bogaert, G. and Bogan, C. and Bohe, A. and Bojtos, P. and Bond, C. and Bondu, F. and Bonnand, R. and Boom, B. A. and Bork, R. and Boschi, V. and Bose, S. and Bouffanais, Y. and Bozzi, A. and Bradaschia, C. and Brady, P. R. and Braginsky, V. B. and Branchesi, M. and Brau, J. E. and Briant, T. and Brillet, A. and Brinkmann, M. and Brisson, V. and Brockill, P. and Brooks, A. F. and Brown, D. A. and Brown, D. D. and Brown, N. M. and Buchanan, C. C. and Buikema, A. and Bulik, T. and Bulten, H. J. and Buonanno, A. and Buskulic, D. and Buy, C. and Byer, R. L. and Cabero, M. and Cadonati, L. and Cagnoli, G. and Cahillane, C. and Bustillo, J. Calder\'on and Callister, T. and Calloni, E. and Camp, J. B. and Cannon, K. C. and Cao, J. and Capano, C. D. and Capocasa, E. and Carbognani, F. and Caride, S. and Diaz, J. Casanueva and Casentini, C. and Caudill, S. and Cavagli\`a, M. and Cavalier, F. and Cavalieri, R. and Cella, G. and Cepeda, C. B. and Baiardi, L. Cerboni and Cerretani, G. and Cesarini, E. and Chakraborty, R. and Chalermsongsak, T. and Chamberlin, S. J. and Chan, M. and Chao, S. and Charlton, P. and Chassande-Mottin, E. and Chen, H. Y. and Chen, Y. and Cheng, C. and Chincarini, A. and Chiummo, A. and Cho, H. S. and Cho, M. and Chow, J. H. and Christensen, N. and Chu, Q. and Chua, S. and Chung, S. and Ciani, G. and Clara, F. and Clark, J. A. and Cleva, F. and Coccia, E. and Cohadon, P.-F. and Colla, A. and Collette, C. G. and Cominsky, L. and Constancio, M. and Conte, A. and Conti, L. and Cook, D. and Corbitt, T. R. and Cornish, N. and Corsi, A. and Cortese, S. and Costa, C. A. and Coughlin, M. W. and Coughlin, S. B. and Coulon, J.-P. and Countryman, S. T. and Couvares, P. and Cowan, E. E. and Coward, D. M. and Cowart, M. J. and Coyne, D. C. and Coyne, R. and Craig, K. and Creighton, J. D. E. and Creighton, T. D. and Cripe, J. and Crowder, S. G. and Cruise, A. M. and Cumming, A. and Cunningham, L. and Cuoco, E. and Canton, T. Dal and Danilishin, S. L. and D'Antonio, S. and Danzmann, K. and Darman, N. S. and Da Silva Costa, C. F. and Dattilo, V. and Dave, I. and Daveloza, H. P. and Davier, M. and Davies, G. S. and Daw, E. J. and Day, R. and De, S. and DeBra, D. and Debreczeni, G. and Degallaix, J. and De Laurentis, M. and Del\'eglise, S. and Del Pozzo, W. and Denker, T. and Dent, T. and Dereli, H. and Dergachev, V. and DeRosa, R. T. and De Rosa, R. and DeSalvo, R. and Dhurandhar, S. and D\'{\i}az, M. C. and Di Fiore, L. and Di Giovanni, M. and Di Lieto, A. and Di Pace, S. and Di Palma, I. and Di Virgilio, A. and Dojcinoski, G. and Dolique, V. and Donovan, F. and Dooley, K. L. and Doravari, S. and Douglas, R. and Downes, T. P. and Drago, M. and Drever, R. W. P. and Driggers, J. C. and Du, Z. and Ducrot, M. and Dwyer, S. E. and Edo, T. B. and Edwards, M. C. and Effler, A. and Eggenstein, H.-B. and Ehrens, P. and Eichholz, J. and Eikenberry, S. S. and Engels, W. and Essick, R. C. and Etzel, T. and Evans, M. and Evans, T. M. and Everett, R. and Factourovich, M. and Fafone, V. and Fair, H. and Fairhurst, S. and Fan, X. and Fang, Q. and Farinon, S. and Farr, B. and Farr, W. M. and Favata, M. and Fays, M. and Fehrmann, H. and Fejer, M. M. and Feldbaum, D. and Ferrante, I. and Ferreira, E. C. and Ferrini, F. and Fidecaro, F. and Finn, L. S. and Fiori, I. and Fiorucci, D. and Fisher, R. P. and Flaminio, R. and Fletcher, M. and Fong, H. and Fournier, J.-D. and Franco, S. and Frasca, S. and Frasconi, F. and Frede, M. and Frei, Z. and Freise, A. and Frey, R. and Frey, V. and Fricke, T. T. and Fritschel, P. and Frolov, V. V. and Fulda, P. and Fyffe, M. and Gabbard, H. A. G. and Gair, J. R. and Gammaitoni, L. and Gaonkar, S. G. and Garufi, F. and Gatto, A. and Gaur, G. and Gehrels, N. and Gemme, G. and Gendre, B. and Genin, E. and Gennai, A. and George, J. and Gergely, L. and Germain, V. and Ghosh, Abhirup and Ghosh, Archisman and Ghosh, S. and Giaime, J. A. and Giardina, K. D. and Giazotto, A. and Gill, K. and Glaefke, A. and Gleason, J. R. and Goetz, E. and Goetz, R. and Gondan, L. and Gonz\'alez, G. and Castro, J. M. Gonzalez and Gopakumar, A. and Gordon, N. A. and Gorodetsky, M. L. and Gossan, S. E. and Gosselin, M. and Gouaty, R. and Graef, C. and Graff, P. B. and Granata, M. and Grant, A. and Gras, S. and Gray, C. and Greco, G. and Green, A. C. and Greenhalgh, R. J. S. and Groot, P. and Grote, H. and Grunewald, S. and Guidi, G. M. and Guo, X. and Gupta, A. and Gupta, M. K. and Gushwa, K. E. and Gustafson, E. K. and Gustafson, R. and Hacker, J. J. and Hall, B. R. and Hall, E. D. and Hammond, G. and Haney, M. and Hanke, M. M. and Hanks, J. and Hanna, C. and Hannam, M. D. and Hanson, J. and Hardwick, T. and Harms, J. and Harry, G. M. and Harry, I. W. and Hart, M. J. and Hartman, M. T. and Haster, C.-J. and Haughian, K. and Healy, J. and Heefner, J. and Heidmann, A. and Heintze, M. C. and Heinzel, G. and Heitmann, H. and Hello, P. and Hemming, G. and Hendry, M. and Heng, I. S. and Hennig, J. and Heptonstall, A. W. and Heurs, M. and Hild, S. and Hoak, D. and Hodge, K. A. and Hofman, D. and Hollitt, S. E. and Holt, K. and Holz, D. E. and Hopkins, P. and Hosken, D. J. and Hough, J. and Houston, E. A. and Howell, E. J. and Hu, Y. M. and Huang, S. and Huerta, E. A. and Huet, D. and Hughey, B. and Husa, S. and Huttner, S. H. and Huynh-Dinh, T. and Idrisy, A. and Indik, N. and Ingram, D. R. and Inta, R. and Isa, H. N. and Isac, J.-M. and Isi, M. and Islas, G. and Isogai, T. and Iyer, B. R. and Izumi, K. and Jacobson, M. B. and Jacqmin, T. and Jang, H. and Jani, K. and Jaranowski, P. and Jawahar, S. and Jim\'enez-Forteza, F. and Johnson, W. W. and Johnson-McDaniel, N. K. and Jones, D. I. and Jones, R. and Jonker, R. J. G. and Ju, L. and Haris, K. and Kalaghatgi, C. V. and Kalogera, V. and Kandhasamy, S. and Kang, G. and Kanner, J. B. and Karki, S. and Kasprzack, M. and Katsavounidis, E. and Katzman, W. and Kaufer, S. and Kaur, T. and Kawabe, K. and Kawazoe, F. and K\'ef\'elian, F. and Kehl, M. S. and Keitel, D. and Kelley, D. B. and Kells, W. and Kennedy, R. and Keppel, D. G. and Key, J. S. and Khalaidovski, A. and Khalili, F. Y. and Khan, I. and Khan, S. and Khan, Z. and Khazanov, E. A. and Kijbunchoo, N. and Kim, C. and Kim, J. and Kim, K. and Kim, Nam-Gyu and Kim, Namjun and Kim, Y.-M. and King, E. J. and King, P. J. and Kinzel, D. L. and Kissel, J. S. and Kleybolte, L. and Klimenko, S. and Koehlenbeck, S. M. and Kokeyama, K. and Koley, S. and Kondrashov, V. and Kontos, A. and Koranda, S. and Korobko, M. and Korth, W. Z. and Kowalska, I. and Kozak, D. B. and Kringel, V. and Krishnan, B. and Kr\'olak, A. and Krueger, C. and Kuehn, G. and Kumar, P. and Kumar, R. and Kuo, L. and Kutynia, A. and Kwee, P. and Lackey, B. D. and Landry, M. and Lange, J. and Lantz, B. and Lasky, P. D. and Lazzarini, A. and Lazzaro, C. and Leaci, P. and Leavey, S. and Lebigot, E. O. and Lee, C. H. and Lee, H. K. and Lee, H. M. and Lee, K. and Lenon, A. and Leonardi, M. and Leong, J. R. and Leroy, N. and Letendre, N. and Levin, Y. and Levine, B. M. and Li, T. G. F. and Libson, A. and Littenberg, T. B. and Lockerbie, N. A. and Logue, J. and Lombardi, A. L. and London, L. T. and Lord, J. E. and Lorenzini, M. and Loriette, V. and Lormand, M. and Losurdo, G. and Lough, J. D. and Lousto, C. O. and Lovelace, G. and L\"uck, H. and Lundgren, A. P. and Luo, J. and Lynch, R. and Ma, Y. and MacDonald, T. and Machenschalk, B. and MacInnis, M. and Macleod, D. M. and Maga\~na-Sandoval, F. and Magee, R. M. and Mageswaran, M. and Majorana, E. and Maksimovic, I. and Malvezzi, V. and Man, N. and Mandel, I. and Mandic, V. and Mangano, V. and Mansell, G. L. and Manske, M. and Mantovani, M. and Marchesoni, F. and Marion, F. and M\'arka, S. and M\'arka, Z. and Markosyan, A. S. and Maros, E. and Martelli, F. and Martellini, L. and Martin, I. W. and Martin, R. M. and Martynov, D. V. and Marx, J. N. and Mason, K. and Masserot, A. and Massinger, T. J. and Masso-Reid, M. and Matichard, F. and Matone, L. and Mavalvala, N. and Mazumder, N. and Mazzolo, G. and McCarthy, R. and McClelland, D. E. and McCormick, S. and McGuire, S. C. and McIntyre, G. and McIver, J. and McManus, D. J. and McWilliams, S. T. and Meacher, D. and Meadors, G. D. and Meidam, J. and Melatos, A. and Mendell, G. and Mendoza-Gandara, D. and Mercer, R. A. and Merilh, E. and Merzougui, M. and Meshkov, S. and Messenger, C. and Messick, C. and Meyers, P. M. and Mezzani, F. and Miao, H. and Michel, C. and Middleton, H. and Mikhailov, E. E. and Milano, L. and Miller, J. and Millhouse, M. and Minenkov, Y. and Ming, J. and Mirshekari, S. and Mishra, C. and Mitra, S. and Mitrofanov, V. P. and Mitselmakher, G. and Mittleman, R. and Moggi, A. and Mohan, M. and Mohapatra, S. R. P. and Montani, M. and Moore, B. C. and Moore, C. J. and Moraru, D. and Moreno, G. and Morriss, S. R. and Mossavi, K. and Mours, B. and Mow-Lowry, C. M. and Mueller, C. L. and Mueller, G. and Muir, A. W. and Mukherjee, Arunava and Mukherjee, D. and Mukherjee, S. and Mukund, N. and Mullavey, A. and Munch, J. and Murphy, D. J. and Murray, P. G. and Mytidis, A. and Nardecchia, I. and Naticchioni, L. and Nayak, R. K. and Necula, V. and Nedkova, K. and Nelemans, G. and Neri, M. and Neunzert, A. and Newton, G. and Nguyen, T. T. and Nielsen, A. B. and Nissanke, S. and Nitz, A. and Nocera, F. and Nolting, D. and Normandin, M. E. N. and Nuttall, L. K. and Oberling, J. and Ochsner, E. and O'Dell, J. and Oelker, E. and Ogin, G. H. and Oh, J. J. and Oh, S. H. and Ohme, F. and Oliver, M. and Oppermann, P. and Oram, Richard J. and O'Reilly, B. and O'Shaughnessy, R. and Ott, C. D. and Ottaway, D. J. and Ottens, R. S. and Overmier, H. and Owen, B. J. and Pai, A. and Pai, S. A. and Palamos, J. R. and Palashov, O. and Palomba, C. and Pal-Singh, A. and Pan, H. and Pan, Y. and Pankow, C. and Pannarale, F. and Pant, B. C. and Paoletti, F. and Paoli, A. and Papa, M. A. and Paris, H. R. and Parker, W. and Pascucci, D. and Pasqualetti, A. and Passaquieti, R. and Passuello, D. and Patricelli, B. and Patrick, Z. and Pearlstone, B. L. and Pedraza, M. and Pedurand, R. and Pekowsky, L. and Pele, A. and Penn, S. and Perreca, A. and Pfeiffer, H. P. and Phelps, M. and Piccinni, O. and Pichot, M. and Pickenpack, M. and Piergiovanni, F. and Pierro, V. and Pillant, G. and Pinard, L. and Pinto, I. M. and Pitkin, M. and Poeld, J. H. and Poggiani, R. and Popolizio, P. and Post, A. and Powell, J. and Prasad, J. and Predoi, V. and Premachandra, S. S. and Prestegard, T. and Price, L. R. and Prijatelj, M. and Principe, M. and Privitera, S. and Prix, R. and Prodi, G. A. and Prokhorov, L. and Puncken, O. and Punturo, M. and Puppo, P. and P\"urrer, M. and Qi, H. and Qin, J. and Quetschke, V. and Quintero, E. A. and Quitzow-James, R. and Raab, F. J. and Rabeling, D. S. and Radkins, H. and Raffai, P. and Raja, S. and Rakhmanov, M. and Ramet, C. R. and Rapagnani, P. and Raymond, V. and Razzano, M. and Re, V. and Read, J. and Reed, C. M. and Regimbau, T. and Rei, L. and Reid, S. and Reitze, D. H. and Rew, H. and Reyes, S. D. and Ricci, F. and Riles, K. and Robertson, N. A. and Robie, R. and Robinet, F. and Rocchi, A. and Rolland, L. and Rollins, J. G. and Roma, V. J. and Romano, J. D. and Romano, R. and Romanov, G. and Romie, J. H. and Rosi\ifmmode \acute{n}\else \'{n}\fi{}ska, D. and Rowan, S. and R\"udiger, A. and Ruggi, P. and Ryan, K. and Sachdev, S. and Sadecki, T. and Sadeghian, L. and Salconi, L. and Saleem, M. and Salemi, F. and Samajdar, A. and Sammut, L. and Sampson, L. M. and Sanchez, E. J. and Sandberg, V. and Sandeen, B. and Sanders, G. H. and Sanders, J. R. and Sassolas, B. and Sathyaprakash, B. S. and Saulson, P. R. and Sauter, O. and Savage, R. L. and Sawadsky, A. and Schale, P. and Schilling, R. and Schmidt, J. and Schmidt, P. and Schnabel, R. and Schofield, R. M. S. and Sch\"onbeck, A. and Schreiber, E. and Schuette, D. and Schutz, B. F. and Scott, J. and Scott, S. M. and Sellers, D. and Sengupta, A. S. and Sentenac, D. and Sequino, V. and Sergeev, A. and Serna, G. and Setyawati, Y. and Sevigny, A. and Shaddock, D. A. and Shaffer, T. and Shah, S. and Shahriar, M. S. and Shaltev, M. and Shao, Z. and Shapiro, B. and Shawhan, P. and Sheperd, A. and Shoemaker, D. H. and Shoemaker, D. M. and Siellez, K. and Siemens, X. and Sigg, D. and Silva, A. D. and Simakov, D. and Singer, A. and Singer, L. P. and Singh, A. and Singh, R. and Singhal, A. and Sintes, A. M. and Slagmolen, B. J. J. and Smith, J. R. and Smith, M. R. and Smith, N. D. and Smith, R. J. E. and Son, E. J. and Sorazu, B. and Sorrentino, F. and Souradeep, T. and Srivastava, A. K. and Staley, A. and Steinke, M. and Steinlechner, J. and Steinlechner, S. and Steinmeyer, D. and Stephens, B. C. and Stevenson, S. P. and Stone, R. and Strain, K. A. and Straniero, N. and Stratta, G. and Strauss, N. A. and Strigin, S. and Sturani, R. and Stuver, A. L. and Summerscales, T. Z. and Sun, L. and Sutton, P. J. and Swinkels, B. L. and Szczepa\ifmmode \acute{n}\else \'{n}\fi{}czyk, M. J. and Tacca, M. and Talukder, D. and Tanner, D. B. and T\'apai, M. and Tarabrin, S. P. and Taracchini, A. and Taylor, R. and Theeg, T. and Thirugnanasambandam, M. P. and Thomas, E. G. and Thomas, M. and Thomas, P. and Thorne, K. A. and Thorne, K. S. and Thrane, E. and Tiwari, S. and Tiwari, V. and Tokmakov, K. V. and Tomlinson, C. and Tonelli, M. and Torres, C. V. and Torrie, C. I. and T\"oyr\"a, D. and Travasso, F. and Traylor, G. and Trifir\`o, D. and Tringali, M. C. and Trozzo, L. and Tse, M. and Turconi, M. and Tuyenbayev, D. and Ugolini, D. and Unnikrishnan, C. S. and Urban, A. L. and Usman, S. A. and Vahlbruch, H. and Vajente, G. and Valdes, G. and Vallisneri, M. and van Bakel, N. and van Beuzekom, M. and van den Brand, J. F. J. and Van Den Broeck, C. and Vander-Hyde, D. C. and van der Schaaf, L. and van Heijningen, J. V. and van Veggel, A. A. and Vardaro, M. and Vass, S. and Vas\'uth, M. and Vaulin, R. and Vecchio, A. and Vedovato, G. and Veitch, J. and Veitch, P. J. and Venkateswara, K. and Verkindt, D. and Vetrano, F. and Vicer\'e, A. and Vinciguerra, S. and Vine, D. J. and Vinet, J.-Y. and Vitale, S. and Vo, T. and Vocca, H. and Vorvick, C. and Voss, D. and Vousden, W. D. and Vyatchanin, S. P. and Wade, A. R. and Wade, L. E. and Wade, M. and Waldman, S. J. and Walker, M. and Wallace, L. and Walsh, S. and Wang, G. and Wang, H. and Wang, M. and Wang, X. and Wang, Y. and Ward, H. and Ward, R. L. and Warner, J. and Was, M. and Weaver, B. and Wei, L.-W. and Weinert, M. and Weinstein, A. J. and Weiss, R. and Welborn, T. and Wen, L. and We\ss{}els, P. and Westphal, T. and Wette, K. and Whelan, J. T. and Whitcomb, S. E. and White, D. J. and Whiting, B. F. and Wiesner, K. and Wilkinson, C. and Willems, P. A. and Williams, L. and Williams, R. D. and Williamson, A. R. and Willis, J. L. and Willke, B. and Wimmer, M. H. and Winkelmann, L. and Winkler, W. and Wipf, C. C. and Wiseman, A. G. and Wittel, H. and Woan, G. and Worden, J. and Wright, J. L. and Wu, G. and Yablon, J. and Yakushin, I. and Yam, W. and Yamamoto, H. and Yancey, C. C. and Yap, M. J. and Yu, H. and Yvert, M. and Zadro\ifmmode \dot{z}\else \.{z}\fi{}ny, A. and Zangrando, L. and Zanolin, M. and Zendri, J.-P. and Zevin, M. and Zhang, F. and Zhang, L. and Zhang, M. and Zhang, Y. and Zhao, C. and Zhou, M. and Zhou, Z. and Zhu, X. J. and Zucker, M. E. and Zuraw, S. E. and Zweizig, J.},
  collaboration = {LIGO Scientific Collaboration and Virgo Collaboration},
  journal = {Phys. Rev. Lett.},
  volume = {116},
  issue = {6},
  pages = {061102},
  numpages = {16},
  year = {2016},
  month = {Feb},
  publisher = {American Physical Society},
  doi = {10.1103/PhysRevLett.116.061102},
  url = {https://link.aps.org/doi/10.1103/PhysRevLett.116.061102}
}

@article{desi2026data,
  title={Data release 1 of the dark energy spectroscopic instrument},
  author={DESI Collaboration and Abdul Karim, M and Adame, AG and Aguado, D and Aguilar, J and Ahlen, S and Alam, S and Aldering, G and Alexander, DM and Alfarsy, R and others},
  journal={The Astronomical Journal},
  volume={171},
  number={5},
  pages={285},
  year={2026},
  publisher={The American Astronomical Society}
}

@article{efstathiou1999cosmic,
  title={Cosmic confusion: degeneracies among cosmological parameters derived from measurements of microwave background anisotropies},
  author={Efstathiou, George and Bond, J Richard},
  journal={Monthly Notices of the Royal Astronomical Society},
  volume={304},
  number={1},
  pages={75--97},
  year={1999},
  publisher={Blackwell Science Ltd Oxford, UK}
}

@article{smith2020massively,
  title={Massively parallel Bayesian inference for transient gravitational-wave astronomy},
  author={Smith, Rory JE and Ashton, Gregory and Vajpeyi, Avi and Talbot, Colm},
  journal={Monthly Notices of the Royal Astronomical Society},
  volume={498},
  number={3},
  pages={4492--4502},
  year={2020},
  publisher={Oxford University Press}
}

@article{Cannon:2011vi,
    author = "Cannon, Kipp and others",
    title = "{Toward Early-Warning Detection of Gravitational Waves from Compact Binary Coalescence}",
    eprint = "1107.2665",
    archivePrefix = "arXiv",
    primaryClass = "astro-ph.IM",
    reportNumber = "LIGO-P0900004, LIGO-P0900004-V34",
    doi = "10.1088/0004-637X/748/2/136",
    journal = "Astrophys. J.",
    volume = "748",
    pages = "136",
    year = "2012"
}

@article{ATLAS:2012yve,
    author = "Aad, Georges and others",
    collaboration = "ATLAS",
    title = "{Observation of a new particle in the search for the Standard Model Higgs boson with the ATLAS detector at the LHC}",
    eprint = "1207.7214",
    archivePrefix = "arXiv",
    primaryClass = "hep-ex",
    reportNumber = "CERN-PH-EP-2012-218",
    doi = "10.1016/j.physletb.2012.08.020",
    journal = "Phys. Lett. B",
    volume = "716",
    pages = "1--29",
    year = "2012"
}

@article{CMS:2012qbp,
    author = "Chatrchyan, Serguei and others",
    collaboration = "CMS",
    title = "{Observation of a New Boson at a Mass of 125 GeV with the CMS Experiment at the LHC}",
    eprint = "1207.7235",
    archivePrefix = "arXiv",
    primaryClass = "hep-ex",
    reportNumber = "CMS-HIG-12-028, CERN-PH-EP-2012-220",
    doi = "10.1016/j.physletb.2012.08.021",
    journal = "Phys. Lett. B",
    volume = "716",
    pages = "30--61",
    year = "2012"
}

@article{khachatryan2017cms,
  title={The CMS trigger system},
  author={Khachatryan, Vardan and Sirunyan, Albert M and Tumasyan, Armen and Adam, Wolfgang and Asilar, E and Bergauer, Thomas and Brandstetter, Johannes and Brondolin, Erica and Dragicevic, Marko and Er{\"o}, Janos and others},
  journal={Journal of Instrumentation},
  volume={12},
  number={01},
  pages={P01020--P01020},
  year={2017}
}

@article{atlas2020operation,
  title={Operation of the ATLAS trigger system in Run 2},
  author={Atlas Collaboration and others},
  journal={Journal of Instrumentation},
  volume={15},
  number={10},
  pages={P10004--P10004},
  year={2020}
}

@article{arnaud2019precision,
  title={Precision laser-based measurements of the single electron response of SPCs for the NEWS-G light dark matter search experiment},
  author={Arnaud, Q and Bard, J-P and Brossard, A and Chapellier, M and Clark, M and Crawford, S and Corcoran, EC and Dastgheibi-Fard, A and Dering, K and Di Stefano, P and others},
  journal={arXiv preprint arXiv:1902.08960},
  year={2019}
}

@article{abdelhameed2019first,
  title={First results from the CRESST-III low-mass dark matter program},
  author={Abdelhameed, Ahmed H and Angloher, G and Bauer, P and Bento, A and Bertoldo, E and Bucci, C and Canonica, L and D’Addabbo, Antonio and Defay, X and Di Lorenzo, S and others},
  journal={Physical Review D},
  volume={100},
  number={10},
  pages={102002},
  year={2019},
  publisher={APS}
}

@techreport{de2017fast,
  title={Fast neural-net based fake track rejection in the LHCb reconstruction},
  author={De Cian, Michel and Stahl, Sascha and Seyfert, Paul and Farry, Stephen},
  year={2017}
}

@article{duarte2018fast,
  title={Fast inference of deep neural networks in FPGAs for particle physics},
  author={Duarte, Javier and Han, Song and Harris, Philip and Jindariani, Sergo and Kreinar, Edward and Kreis, Benjamin and Ngadiuba, Jennifer and Pierini, Maurizio and Rivera, Ryan and Tran, Nhan and others},
  journal={Journal of instrumentation},
  volume={13},
  number={07},
  pages={P07027},
  year={2018},
  publisher={IOP Publishing}
}

@article{schaefer2025advancing,
  title={Advancing the CMS Level-1 Trigger: Jet Tagging with DeepSets at the HL-LHC},
  author={Schaefer, Stella and Brown, Christopher and Hoang, Duc and Summers, Sioni and Wuchterl, Sebastian},
  journal={arXiv preprint arXiv:2509.24371},
  year={2025}
}

@article{Govorkova:2021utb,
    author = "Govorkova, Ekaterina and others",
    title = "{Autoencoders on field-programmable gate arrays for real-time, unsupervised new physics detection at 40 MHz at the Large Hadron Collider}",
    eprint = "2108.03986",
    archivePrefix = "arXiv",
    primaryClass = "physics.ins-det",
    reportNumber = "FERMILAB-PUB-21-487-CMS, FERMILAB-PUB-21-487-CMS",
    doi = "10.1038/s42256-022-00441-3",
    journal = "Nature Mach. Intell.",
    volume = "4",
    pages = "154--161",
    year = "2022"
}

@article{george2018deep,
  title={Deep neural networks to enable real-time multimessenger astrophysics},
  author={George, Daniel and Huerta, EA},
  journal={Physical Review D},
  volume={97},
  number={4},
  pages={044039},
  year={2018},
  publisher={APS}
}

@article{raikman2024gwak,
  title={GWAK: gravitational-wave anomalous knowledge with recurrent autoencoders},
  author={Raikman, Ryan and Moreno, Eric A and Govorkova, Ekaterina and Marx, Ethan J and Gunny, Alec and Benoit, William and Chatterjee, Deep and Omer, Rafia and Saleem, Muhammed and Rankin, Dylan S and others},
  journal={Machine Learning: Science and Technology},
  volume={5},
  number={2},
  pages={025020},
  year={2024},
  publisher={IOP Publishing}
}

@article{langer2024probing,
  title={Probing the effects of broken symmetries in machine learning},
  author={Langer, Marcel F and Pozdnyakov, Sergey N and Ceriotti, Michele},
  journal={Machine Learning: Science and Technology},
  volume={5},
  number={4},
  pages={04LT01},
  year={2024},
  publisher={IOP Publishing}
}

@article{nabat2025learning,
  title={Learning broken symmetries with approximate invariance},
  author={Nabat, Seth and Ghosh, Aishik and Witkowski, Edmund and Kasieczka, Gregor and Whiteson, Daniel},
  journal={Physical Review D},
  volume={111},
  number={7},
  pages={072002},
  year={2025},
  publisher={APS}
}

@article{Aarts:2026zzr,
    author = {Aarts, Gert and Habibi, Diaa E. and Ipp, Andreas and M{\"u}ller, David I. and Ranner, Thomas R. and Wang, Lingxiao and Wang, Wei and Zhu, Qianteng},
    title = "{Generalizable Equivariant Diffusion Models for Non-Abelian Lattice Gauge Theory}",
    eprint = "2601.19552",
    archivePrefix = "arXiv",
    primaryClass = "hep-lat",
    reportNumber = "RIKEN-iTHEMS-Report-26",
    month = "1",
    year = "2026"
}

@article{Feickert:2021ajf,
    author = "Feickert, Matthew and Nachman, Benjamin",
    title ="{A Living Review of Machine Learning for Particle Physics}",
    eprint = "2102.02770",
    archivePrefix = "arXiv",
    primaryClass = "hep-ph",
    month = "2",
    year = "2021"
}

@article{Birk:2024knn,
    author = "Birk, Joschka and Hallin, Anna and Kasieczka, Gregor",
    title = "{OmniJet-$\alpha$: The first cross-task foundation model for particle physics}",
    eprint = "2403.05618",
    archivePrefix = "arXiv",
    primaryClass = "hep-ph",
    month = "3",
    year = "2024"
}

@inproceedings{Hallin:2025ywf,
    author = "Hallin, Anna",
    title = "{Foundation models for high-energy physics}",
    booktitle = "{2nd European AI for Fundamental Physics Conference}",
    eprint = "2509.21434",
    archivePrefix = "arXiv",
    primaryClass = "hep-ph",
    month = "9",
    year = "2025"
}

@article{Harris:2024sra,
    author = "Harris, Philip and Kagan, Michael and Krupa, Jeffrey and Maier, Benedikt and Woodward, Nathaniel",
    title = "{Re-Simulation-based Self-Supervised Learning for Pre-Training Foundation Models}",
    eprint = "2403.07066",
    archivePrefix = "arXiv",
    primaryClass = "hep-ph",
    month = "3",
    year = "2024"
}

@article{Golling:2024abg,
    author = "Golling, Tobias and Heinrich, Lukas and Kagan, Michael and Klein, Samuel and Leigh, Matthew and Osadchy, Margarita and Raine, John Andrew",
    title = "{Masked particle modeling on sets: towards self-supervised high energy physics foundation models}",
    eprint = "2401.13537",
    archivePrefix = "arXiv",
    primaryClass = "hep-ph",
    doi = "10.1088/2632-2153/ad64a8",
    journal = "Mach. Learn. Sci. Tech.",
    volume = "5",
    number = "3",
    pages = "035074",
    year = "2024"
}

@article{Vigl:2024lat,
    author = "Vigl, Matthias and Hartman, Nicole and Heinrich, Lukas",
    title = "{Finetuning foundation models for joint analysis optimization in High Energy Physics}",
    eprint = "2401.13536",
    archivePrefix = "arXiv",
    primaryClass = "hep-ex",
    doi = "10.1088/2632-2153/ad55a3",
    journal = "Mach. Learn. Sci. Tech.",
    volume = "5",
    number = "2",
    pages = "025075",
    year = "2024"
}

@article{Mikuni:2025tar,
    author = "Mikuni, Vinicius and Nachman, Benjamin",
    title = "{Method to simultaneously facilitate all jet physics tasks}",
    eprint = "2502.14652",
    archivePrefix = "arXiv",
    primaryClass = "hep-ph",
    doi = "10.1103/PhysRevD.111.054015",
    journal = "Phys. Rev. D",
    volume = "111",
    number = "5",
    pages = "054015",
    year = "2025"
}

@article{Mikuni:2024qsr,
    author = "Mikuni, Vinicius and Nachman, Benjamin",
    title = "{Solving key challenges in collider physics with foundation models}",
    eprint = "2404.16091",
    archivePrefix = "arXiv",
    primaryClass = "hep-ph",
    doi = "10.1103/PhysRevD.111.L051504",
    journal = "Phys. Rev. D",
    volume = "111",
    number = "5",
    pages = "L051504",
    year = "2025"
}

@article{Leigh:2024ked,
    author = "Leigh, Matthew and Klein, Samuel and Charton, Fran{\c{c}}ois and Golling, Tobias and Heinrich, Lukas and Kagan, Michael and Ochoa, In{\^e}s and Osadchy, Margarita",
    title = "{Is Tokenization Needed for Masked Particle Modelling?}",
    eprint = "2409.12589",
    archivePrefix = "arXiv",
    primaryClass = "hep-ph",
    month = "9",
    year = "2024"
}

@article{Bardhan:2025icr,
    author = "Bardhan, Jai and Agrawal, Radhikesh and Tilak, Abhiram and Neeraj, Cyrin and Mitra, Subhadip",
    title = "{HEP-JEPA: A foundation model for collider physics using joint embedding predictive architecture}",
    eprint = "2502.03933",
    archivePrefix = "arXiv",
    primaryClass = "cs.LG",
    month = "2",
    year = "2025"
}

@article{Qu:2022mxj,
    author = "Hayrapetyan, Aram and others",
    collaboration = "CMS",
    title = "{Particle transformers for identifying Lorentz-boosted Higgs bosons decaying to a pair of W bosons}",
    eprint = "2604.09809",
    archivePrefix = "arXiv",
    primaryClass = "hep-ex",
    reportNumber = "CMS-JME-25-001, CERN-EP-2026-111",
    month = "4",
    year = "2026"
}

@article{Mikuni:2025ocp,
    author = "Mikuni, Vinicius and Elsharkawy, Ibrahim and Nachman, Benjamin",
    title = "{OmniCosmos: Transferring Particle Physics Knowledge Across the Cosmos}",
    eprint = "2512.24422",
    archivePrefix = "arXiv",
    primaryClass = "astro-ph.CO",
    month = "12",
    year = "2025"
}

@article{Elsharkawy:2026kwp,
    author = "Elsharkawy, Ibrahim and Mikuni, Vinicius and Bhimji, Wahid and Nachman, Benjamin",
    title = "{OmniMol: Transferring Particle Physics Knowledge to Molecular Dynamics with Point-Edge Transformers}",
    eprint = "2601.10791",
    archivePrefix = "arXiv",
    primaryClass = "physics.chem-ph",
    month = "1",
    year = "2026"
}

@article{Lu:2024ict,
    author = "Lu, Junjian and Liu, Siwei and Kobylianskii, Dmitrii and Dreyer, Etienne and Gross, Eilam and Liang, Shangsong",
    title = "{PASCL: supervised contrastive learning with perturbative augmentation for particle decay reconstruction}",
    eprint = "2402.11538",
    archivePrefix = "arXiv",
    primaryClass = "hep-ph",
    doi = "10.1088/2632-2153/ad8060",
    journal = "Mach. Learn. Sci. Tech.",
    volume = "5",
    number = "4",
    pages = "045028",
    year = "2024"
}

@article{Metzger:2025ecl,
    author = "Metzger, Kyle and Xu, Lana and Sodini, Mia and Arrestad, Thea K. and Govorkova, Katya and Grosso, Gaia and Harris, Philip",
    title = "{Anomaly-preserving contrastive neural embeddings for end-to-end model-independent searches at the LHC}",
    eprint = "2502.15926",
    archivePrefix = "arXiv",
    primaryClass = "hep-ex",
    doi = "10.1103/5n77-ynsp",
    journal = "Phys. Rev. D",
    volume = "112",
    number = "7",
    pages = "072011",
    year = "2025"
}

@article{bright2026autoscidact,
  title={Autoscidact: Automated scientific discovery through contrastive embedding and hypothesis testing},
  author={Bright-Thonney, Sam and Reissel, Christina and Grosso, Gaia and Woodward, Nathaniel and Govorkova, Katya and Novak, Andrzej and Park, Sangeon and Moreno, Eric and Harris, Philip},
  journal={Advances in Neural Information Processing Systems},
  volume={38},
  pages={88575--88614},
  year={2026}
}

@article{Sheldon:2024sbe,
    author = "Sheldon, Liam Rankin and Rankin, Dylan Sheldon and Harris, Philip",
    title = "{MACK: Mismodeling addressed with contrastive knowledge}",
    eprint = "2410.13947",
    archivePrefix = "arXiv",
    primaryClass = "hep-ph",
    doi = "10.21468/SciPostPhys.18.5.150",
    journal = "SciPost Phys.",
    volume = "18",
    number = "5",
    pages = "150",
    year = "2025"
}

@article{CMS-DP-2024-066,
      collaboration = "CMS",
      title         = "{A unified approach for jet tagging in Run 3 at
                       $\sqrt{s}$=13.6 TeV in CMS}",
      year          = "2024",
      url           = "https://cds.cern.ch/record/2904702",
}

@article{ATLAS:2025dkv,
    author = "Aad, Georges and others",
    collaboration = "ATLAS",
    title = "{Transforming jet flavour tagging at ATLAS}",
    eprint = "2505.19689",
    archivePrefix = "arXiv",
    primaryClass = "hep-ex",
    reportNumber = "CERN-EP-2025-103",
    doi = "10.1038/s41467-025-65059-6",
    journal = "Nature Commun.",
    volume = "17",
    number = "1",
    pages = "541",
    year = "2026"
}

@article{Krause:2024avx,
    author = "Amram, Oz and others",
    editor = "Krause, Claudius and Faucci Giannelli, Michele and Kasieczka, Gregor and Nachman, Benjamin and Salamani, Dalila and Shih, David and Zaborowska, Anna",
    title = "{CaloChallenge 2022: a community challenge for fast calorimeter simulation}",
    eprint = "2410.21611",
    archivePrefix = "arXiv",
    primaryClass = "physics.ins-det",
    reportNumber = "HEPHY-ML-24-05, FERMILAB-PUB-24-0728-CMS, TTK-24-43",
    doi = "10.1088/1361-6633/ae1304",
    journal = "Rept. Prog. Phys.",
    volume = "88",
    number = "11",
    pages = "116201",
    year = "2025"
}

@article{anderlini2021machine,
  title={Machine learning for the LHCb simulation},
  author={Anderlini, Lucio},
  journal={arXiv preprint arXiv:2110.07925},
  year={2021}
}

@article{touranakou2022particle,
  title={Particle-based fast jet simulation at the LHC with variational autoencoders},
  author={Touranakou, Mary and Chernyavskaya, Nadezda and Duarte, Javier and Gunopulos, Dimitrios and Kansal, Raghav and Orzari, Breno and Pierini, Maurizio and Tomei, Thiago and Vlimant, Jean-Roch},
  journal={Machine Learning: Science and Technology},
  volume={3},
  number={3},
  pages={035003},
  year={2022},
  publisher={IOP Publishing}
}

@techreport{javurkova2024fast,
  title={The Fast Simulation Program of ATLAS at the LHC},
  author={Javurkova, Martina},
  year={2024},
  institution={ATL-COM-SOFT-2024-032}
}

@article{mazurek2025machine,
  title={Machine learning in LHCb Simulation: From fast to flash},
  author={Mazurek, Micha{\l}},
  journal={arXiv preprint arXiv:2511.02020},
  year={2025}
}

@article{
doi:10.1073/pnas.1821458116,
author = {Siyu He  and Yin Li  and Yu Feng  and Shirley Ho  and Siamak Ravanbakhsh  and Wei Chen  and Barnabás Póczos },
title = {Learning to predict the cosmological structure formation},
journal = {Proceedings of the National Academy of Sciences},
volume = {116},
number = {28},
pages = {13825-13832},
year = {2019},
doi = {10.1073/pnas.1821458116},
URL = {https://www.pnas.org/doi/abs/10.1073/pnas.1821458116},
eprint = {https://www.pnas.org/doi/pdf/10.1073/pnas.1821458116}
}

@article{rodriguez2018fast,
  title={Fast cosmic web simulations with generative adversarial networks},
  author={Rodriguez, Andres C and Kacprzak, Tomasz and Lucchi, Aurelien and Amara, Adam and Sgier, Raphael and Fluri, Janis and Hofmann, Thomas and R{\'e}fr{\'e}gier, Alexandre},
  journal={Computational Astrophysics and Cosmology},
  volume={5},
  number={1},
  pages={1--11},
  year={2018},
  publisher={Springer}
}

@article{10.1088/2632-2153/ae73e3,
	author={Cappelli, Pietro and Grosso, Gaia and Letizia, Marco and Reyes-González, Humberto and Zanetti, Marco},
	title={Learning to validate generative models: a goodness-of-fit approach},
	journal={Machine Learning: Science and Technology},
	url={http://iopscience.iop.org/article/10.1088/2632-2153/ae73e3},
	year={2026},
}

@article{Kansal:2022spb,
    author = "Kansal, Raghav and Li, Anni and Duarte, Javier and Chernyavskaya, Nadezda and Pierini, Maurizio and Orzari, Breno and Tomei, Thiago",
    title = "{Evaluating generative models in high energy physics}",
    eprint = "2211.10295",
    archivePrefix = "arXiv",
    primaryClass = "hep-ex",
    reportNumber = "FERMILAB-PUB-22-872-CMS-PPD",
    doi = "10.1103/PhysRevD.107.076017",
    journal = "Phys. Rev. D",
    volume = "107",
    number = "7",
    pages = "076017",
    year = "2023"
}

@article{kamdar2016machine,
  title={Machine learning and cosmological simulations--ii. hydrodynamical simulations},
  author={Kamdar, Harshil M and Turk, Matthew J and Brunner, Robert J},
  journal={Monthly Notices of the Royal Astronomical Society},
  volume={457},
  number={2},
  pages={1162--1179},
  year={2016},
  publisher={Oxford University Press}
}

@article{cranmer2020frontier,
  title={The frontier of simulation-based inference},
  author={Cranmer, Kyle and Brehmer, Johann and Louppe, Gilles},
  journal={Proceedings of the National Academy of Sciences},
  volume={117},
  number={48},
  pages={30055--30062},
  year={2020},
  publisher={National Academy of Sciences}
}

@article{Astrand_2026,
doi = {10.1088/2632-2153/ae35cc},
url = {https://doi.org/10.1088/2632-2153/ae35cc},
year = {2026},
month = {jan},
publisher = {IOP Publishing},
volume = {7},
number = {1},
pages = {013001},
author = {Astrand, S and Boggia, L and Borsato, M and Bozianu, L and Cocha Toapaxi, C E and Giasemis, F I and Hansen, J and Inkaew, P and Iversen, K E and Jawahar, P and Pineiro Monteagudo, H and Olocco, M and Schramm, S},
title = {Perspective on machine learning for real-time analysis at the Large Hadron Collider experiments ALICE, ATLAS, CMS and LHCb},
journal = {Machine Learning: Science and Technology}
}

@article{finzi2026entropy,
  title={From Entropy to Epiplexity: Rethinking Information for Computationally Bounded Intelligence},
  author={Finzi, Marc and Qiu, Shikai and Jiang, Yiding and Izmailov, Pavel and Kolter, J Zico and Wilson, Andrew Gordon},
  journal={arXiv preprint arXiv:2601.03220},
  year={2026}
}

@article{goldblum2023no,
  title={The no free lunch theorem, kolmogorov complexity, and the role of inductive biases in machine learning},
  author={Goldblum, Micah and Finzi, Marc and Rowan, Keefer and Wilson, Andrew Gordon},
  journal={arXiv preprint arXiv:2304.05366},
  year={2023}
}

@article{wolpert1996lack,
  title={The lack of a priori distinctions between learning algorithms},
  author={Wolpert, David H},
  journal={Neural computation},
  volume={8},
  number={7},
  pages={1341--1390},
  year={1996},
  publisher={MIT Press One Rogers Street, Cambridge, MA 02142-1209, USA journals-info~…}
}

@techreport{wolpert1995no,
  title={No free lunch theorems for search},
  author={Wolpert, David H and Macready, William G and others},
  year={1995},
  institution={Technical Report SFI-TR-95-02-010, Santa Fe Institute}
}

@article{wolpert2002no,
  title={No free lunch theorems for optimization},
  author={Wolpert, David H and Macready, William G},
  journal={IEEE transactions on evolutionary computation},
  volume={1},
  number={1},
  pages={67--82},
  year={2002},
  publisher={IEEE}
}

@article{valle2018deep,
  title={Deep learning generalizes because the parameter-function map is biased towards simple functions},
  author={Valle-Perez, Guillermo and Camargo, Chico Q and Louis, Ard A},
  journal={arXiv preprint arXiv:1805.08522},
  year={2018}
}

@article{huh2021low,
  title={The low-rank simplicity bias in deep networks},
  author={Huh, Minyoung and Mobahi, Hossein and Zhang, Richard and Cheung, Brian and Agrawal, Pulkit and Isola, Phillip},
  journal={arXiv preprint arXiv:2103.10427},
  year={2021}
}

@article{PhysRevD.111.032010,
  title = {Resimulation-based self-supervised learning for pretraining physics foundation models},
  author = {Harris, P. and Krupa, J. and Kagan, M. and Maier, B. and Woodward, N.},
  journal = {Phys. Rev. D},
  volume = {111},
  issue = {3},
  pages = {032010},
  numpages = {14},
  year = {2025},
  month = {Feb},
  publisher = {American Physical Society},
  doi = {10.1103/PhysRevD.111.032010},
  url = {https://link.aps.org/doi/10.1103/PhysRevD.111.032010}
}

@article{5n77-ynsp,
  title = {Anomaly-preserving contrastive neural embeddings for end-to-end model-independent searches at the LHC},
  author = {Metzger, Kyle and Xu, Lana and Sodini, Mia and \AA{}rrestad, Thea K. and Govorkova, Katya and Grosso, Gaia and Harris, Philip},
  journal = {Phys. Rev. D},
  volume = {112},
  issue = {7},
  pages = {072011},
  numpages = {20},
  year = {2025},
  month = {Oct},
  publisher = {American Physical Society},
  doi = {10.1103/5n77-ynsp},
  url = {https://link.aps.org/doi/10.1103/5n77-ynsp}
}

@article{xu2020theory,
  title={A theory of usable information under computational constraints},
  author={Xu, Yilun and Zhao, Shengjia and Song, Jiaming and Stewart, Russell and Ermon, Stefano},
  journal={arXiv preprint arXiv:2002.10689},
  year={2020}
}

@book{longpre1986resource,
  title={RESOURCE BOUNDED KOLMOGOROV COMPLEXITY, A LINK BETWEEN COMPUTATIONAL COMPLEXITY AND INFORMATION THEORY (RANDOMNESS)},
  author={Longpr{\'e}, Luc},
  year={1986},
  publisher={Cornell University}
}

@article{allender2006power,
  title={Power from random strings},
  author={Allender, Eric and Buhrman, Harry and Kouck{\`y}, Michal and Van Melkebeek, Dieter and Ronneburger, Detlef},
  journal={SIAM Journal on Computing},
  volume={35},
  number={6},
  pages={1467--1493},
  year={2006},
  publisher={SIAM}
}

@software{getphysicsdone,
  author = {{Physical Superintelligence PBC}},
  title = {Get Physics Done (GPD)},
  version = {1.2.2},
  year = {2026},
  url = {https://github.com/psi-oss/get-physics-done},
  license = {Apache-2.0}
}

@misc{getanalysisdone,
  author = {Moreno, Eric and others},
  title = {Get Analysis Done: Autonomous High-Energy Physics Analysis with LLMs},
  url = {https://github.com/eric-moreno/get-analysis-done},
  year = {2024}
}

@misc{autoresearch,
  author = {Karpathy, Andrej},
  title = {Autoresearch},
  url = {https://github.com/karpathy/autoresearch},
  year = {2024}
}

@article{dorigo2023toward,
  title={Toward the end-to-end optimization of particle physics instruments with differentiable programming},
  author={Dorigo, Tommaso and Giammanco, Andrea and Vischia, Pietro and Aehle, Max and Bawaj, Mateusz and Boldyrev, Alexey and de Castro Manzano, Pablo and Derkach, Denis and Donini, Julien and Edelen, Auralee and others},
  journal={Reviews in Physics},
  volume={10},
  pages={100085},
  year={2023},
  publisher={Elsevier}
}

@article{aehle2025progress,
  title={Progress in end-to-end optimization of fundamental physics experimental apparata with differentiable programming},
  author={Aehle, Max and Arsini, Lorenzo and Barreiro, R Bel{\'e}n and Belias, Anastasios and Boldyrev, Alexey and Bury, Florian and Cebrian, Susana and Demin, Alexander and Dickinson, Jennet and Donini, Julien and others},
  journal={Reviews in physics},
  volume={13},
  pages={100120},
  year={2025},
  publisher={Elsevier}
}

@article{miromind2026mirothinker,
  title={MiroThinker-1.7 \& H1: Towards Heavy-Duty Research Agents via Verification},
  author={MiroMind Team and Bai, S. and Bing, L. and Lei, L. and Li, R. and Li, X. and Lin, X. and Min, E. and Su, L. and Wang, B. and Wang, L. and Wang, L. and Wang, S. and Wang, X. and Zhang, Y. and Zhang, Z. and others},
  journal={arXiv preprint arXiv:2603.15726},
  year={2026}
}

@article{Moreno:2026mqk,
    author = "Moreno, Eric A. and Bright-Thonney, Samuel and Novak, Andrzej and Garcia, Dolores and Harris, Philip",
    title = "{AI Agents Can Already Autonomously Perform Experimental High Energy Physics}",
    eprint = "2603.20179",
    archivePrefix = "arXiv",
    primaryClass = "hep-ex",
    month = "3",
    year = "2026"
}

@article{badea2026agentic,
  title={Agentic AI--Physicist Collaboration in Experimental Particle Physics: A Proof-of-Concept Measurement with LEP Open Data},
  author={Badea, Anthony and Chen, Yi and Lee, Yen-Jie},
  journal={arXiv preprint arXiv:2603.05735},
  year={2026}
}

@article{lu2024ai,
  title={The ai scientist: Towards fully automated open-ended scientific discovery},
  author={Lu, Chris and Lu, Cong and Lange, Robert Tjarko and Foerster, Jakob and Clune, Jeff and Ha, David},
  journal={arXiv preprint arXiv:2408.06292},
  year={2024}
}

@article{mcgreivy2025seeing,
  title={Seeing the Forest Through the Trees: Knowledge Retrieval for Streamlining Particle Physics Analysis},
  author={McGreivy, James and Delaney, Blaise and Beck, Anja and Williams, Mike},
  journal={arXiv preprint arXiv:2509.06855},
  year={2025}
}

@article{menzo2025heptapod,
  title={HEPTAPOD: Orchestrating High Energy Physics Workflows Towards Autonomous Agency},
  author={Menzo, Tony and Roman, Alexander and Gleyzer, Sergei and Matchev, Konstantin and Fleming, George T and H{\"o}che, Stefan and Mrenna, Stephen and Shyamsundar, Prasanth},
  journal={arXiv preprint arXiv:2512.15867},
  year={2025}
}

@article{diefenbacher2025agents,
  title={Agents of Discovery},
  author={Diefenbacher, Sascha and Hallin, Anna and Kasieczka, Gregor and Kr{\"a}mer, Michael and Lauscher, Anne and Lukas, Tim},
  journal={arXiv preprint arXiv:2509.08535},
  year={2025}
}

@article{goodfellow2014explaining,
  title={Explaining and harnessing adversarial examples},
  author={Goodfellow, Ian J and Shlens, Jonathon and Szegedy, Christian},
  journal={arXiv preprint arXiv:1412.6572},
  year={2014}
}

@inproceedings{moura2021lean,
  title={The lean 4 theorem prover and programming language},
  author={Moura, Leonardo de and Ullrich, Sebastian},
  booktitle={International Conference on Automated Deduction},
  pages={625--635},
  year={2021},
  organization={Springer}
}

@article{christiano2017deep,
  title={Deep reinforcement learning from human preferences},
  author={Christiano, Paul F and Leike, Jan and Brown, Tom and Martic, Miljan and Legg, Shane and Amodei, Dario},
  journal={Advances in neural information processing systems},
  volume={30},
  year={2017}
}

@article{ouyang2022training,
  title={Training language models to follow instructions with human feedback},
  author={Ouyang, Long and Wu, Jeffrey and Jiang, Xu and Almeida, Diogo and Wainwright, Carroll and Mishkin, Pamela and Zhang, Chong and Agarwal, Sandhini and Slama, Katarina and Ray, Alex and others},
  journal={Advances in neural information processing systems},
  volume={35},
  pages={27730--27744},
  year={2022}
}

@inproceedings{lightman2023let,
  title={Let's verify step by step},
  author={Lightman, Hunter and Kosaraju, Vineet and Burda, Yuri and Edwards, Harrison and Baker, Bowen and Lee, Teddy and Leike, Jan and Schulman, John and Sutskever, Ilya and Cobbe, Karl},
  booktitle={The twelfth international conference on learning representations},
  year={2023}
}

@article{bai2022constitutional,
  title={Constitutional ai: Harmlessness from ai feedback},
  author={Bai, Yuntao and Kadavath, Saurav and Kundu, Sandipan and Askell, Amanda and Kernion, Jackson and Jones, Andy and Chen, Anna and Goldie, Anna and Mirhoseini, Azalia and McKinnon, Cameron and others},
  journal={arXiv preprint arXiv:2212.08073},
  year={2022}
}

@article{uesato2022solving,
  title={Solving math word problems with process-and outcome-based feedback},
  author={Uesato, Jonathan and Kushman, Nate and Kumar, Ramana and Song, Francis and Siegel, Noah and Wang, Lisa and Creswell, Antonia and Irving, Geoffrey and Higgins, Irina},
  journal={arXiv preprint arXiv:2211.14275},
  year={2022}
}

@book{cover1999elements,
  title={Elements of information theory},
  author={Cover, Thomas M},
  year={1999},
  publisher={John Wiley \& Sons}
}

@unpublished{ThaisManuscript-THAAFS-2,
	author = {Savannah Thais and Roberto Trotta and Nathan Suri and Emily Sullivan and Viyan Poonamallee and Tanaporn Na Narong and Rupert Croft and Nicole Hartman},
	title = {Ai for Science Needs Scientific Alignment},
	year = {manuscript}
}

@article{Thais2024,
  author  = {Thais, Savannah},
  title   = {Physics and the empirical gap of trustworthy {AI}},
  journal = {Nature Reviews Physics},
  volume  = {6},
  number  = {11},
  pages   = {640--641},
  year    = {2024},
  month   = nov,
  doi     = {10.1038/s42254-024-00772-7},
  url     = {https://doi.org/10.1038/s42254-024-00772-7},
  issn    = {2522-5820},
  publisher = {Springer Nature}
}

@article{simon2026there,
  title={There will be a scientific theory of deep learning},
  author={Simon, Jamie and Kunin, Daniel and Atanasov, Alexander and Boix-Adser{\`a}, Enric and Bordelon, Blake and Cohen, Jeremy and Ghosh, Nikhil and Guth, Florentin and Jacot, Arthur and Kamb, Mason and others},
  journal={arXiv preprint arXiv:2604.21691},
  year={2026}
}

@article{haussmann2026uncertainty,
  title={Uncertainty in Physics and AI: Taxonomy, Quantification, and Validation},
  author={Hau{\ss}mann, Manuel and Winterhalder, Ramon and Ubiali, Maria},
  journal={arXiv preprint arXiv:2605.10378},
  year={2026}
}

@article{Gambhir2026,
  author        = {Gambhir, Rikab and Lucie-Smith, Luisa and Thaler, Jesse},
  title         = {Interpreting ``Interpretability'' and Explaining ``Explainability'' in Machine Learning in Physics},
  year          = {2026},
  eprint        = {2606.26228},
  archivePrefix = {arXiv},
  primaryClass  = {physics.data-an},
  note          = {31 pages, 3 figures. Part of the VERaiPHY Initiative. MIT-CTP/6058}
}

@article{spinner2026symmetry,
  title={Symmetry-Informed Machine Learning for Fundamental Physics},
  author={Spinner, Jonas and Villar, Soledad},
  journal={in preparation},
  year={2026}
}

@article{max2026simulation,
  title={Symmetry-Informed Machine Learning for Fundamental Physics},
  author={Dax, Max and Heimel, Theo and Louppe, Gilles},
  journal={in preparation},
  year={2026}
}

@article{cruz2026unknown,
  title={Unknown Unknowns in Machine Learning for Physics},
  author={Cruz-Martinez, Jan and Cuesta-Lazaro, Carolina and Held, Alexander and Kagan Michael},
  journal={in preparation},
  year={2026}
}

@article{klein2026representation,
  title={Representation Learning in Fundamental Physics},
  author={Klein, Samuel and Muthukrishna2, Daniel and Scaife, Anna},
  journal={in preparation},
  year={2026}
}

@article{lavie2026statistical,
  title={Statistical Properties of Training \& Generalization},
  author={Lavie, Itay and Levi, Noam and Kahn, Yonatan},
  journal={arXiv preprint arXiv:2606.20299},
  year={2026}
}

@article{diefenbacher2026generative,
  title={Generative Models and Statistical Validation},
  author={Diefenbacher, Sascha and Schweitzer, Sofia Palacios and Kasieczka, Gregor},
  journal={arXiv preprint arXiv:2605.30453},
  year={2026}
}

@article{amram2026model,
  title={Model-Agnostic Signal Discovery with Machine Learning: Bridging the Gap Between Theory and Practice},
  author={Amram, Oz and Letizia, Marco and Kuusela, Mikael},
  journal={arXiv preprint arXiv:2605.31103},
  year={2026}
}

@article{klein2005blind,
  title={Blind analysis in nuclear and particle physics},
  author={Klein, Joshua R and Roodman, Aaron},
  journal={Annu. Rev. Nucl. Part. Sci.},
  volume={55},
  number={1},
  pages={141--163},
  year={2005},
  publisher={Annual Reviews}
}

@article{agostinelli2003geant4,
  title={Geant4—a simulation toolkit},
  author={Agostinelli, Sea and Allison, John and Amako, K al and Apostolakis, John and Araujo, Henrique and Arce, Pedro and Asai, Makoto and Axen, D and Banerjee, Swagato and Barrand, GJNI and others},
  journal={Nuclear instruments and methods in physics research section A: Accelerators, Spectrometers, Detectors and Associated Equipment},
  volume={506},
  number={3},
  pages={250--303},
  year={2003},
  publisher={Elsevier}
}

@article{abbott2023gwtc,
  title={GWTC-3: Compact binary coalescences observed by LIGO and Virgo during the second part of the third observing run},
  author={Abbott, Richard and Abbott, Thomas D and Acernese, Fausto and Ackley, Kimberly and Adams, Carl and Adhikari, Naresh and Adhikari, Radha X and Adya, Vaishali B and Affeldt, Christoph and Agarwal, Deepali and others},
  journal={Physical Review X},
  volume={13},
  number={4},
  pages={041039},
  year={2023},
  publisher={APS}
}

@article{abac2026gwtc,
  title={GWTC-4.0: Updating the gravitational-wave transient catalog with observations from the first part of the fourth LIGO--Virgo--KAGRA observing run},
  author={Abac, AG and Abouelfettouh, I and Acernese, F and Ackley, K and Adamcewicz, C and Adhicary, S and Adhikari, D and Adhikari, N and Adhikari, RX and Adkins, VK and others},
  journal={The Astrophysical Journal Letters},
  volume={1004},
  number={2},
  pages={L22},
  year={2026},
  publisher={The American Astronomical Society}
}

@article{evans2008lhc,
  title={LHC machine},
  author={Evans, Lyndon and Bryant, Philip},
  journal={Journal of instrumentation},
  volume={3},
  number={08},
  pages={S08001--S08001},
  year={2008}
}

@article{aprile2018dark,
  title={Dark matter search results from a one ton-year exposure of XENON1T},
  author={Aprile, Elena and Aalbers, J and Agostini, F and Alfonsi, M and Althueser, L and Amaro, FD and Anthony, M and Arneodo, F and Baudis, L and Bauermeister, Boris and others},
  journal={Physical review letters},
  volume={121},
  number={11},
  pages={111302},
  year={2018},
  publisher={APS}
}

@article{davies1977,
  title={Hypothesis testing when a nuisance parameter is present only under the alternative},
  author={Davies, Robert B},
  journal={Biometrika},
  volume={64},
  number={2},
  pages={247--254},
  year={1977},
  publisher={Oxford University Press}
}

@article{davies1987,
 ISSN = {00063444},
 URL = {http://www.jstor.org/stable/2336019},
 author = {Robert B. Davies},
 journal = {Biometrika},
 number = {1},
 pages = {33--43},
 publisher = {[Oxford University Press, Biometrika Trust]},
 title = {Hypothesis Testing when a Nuisance Parameter is Present Only Under the Alternatives},
 urldate = {2026-03-22},
 volume = {74},
 year = {1987}
}

@article{Vitells:2011da,
    author = "Vitells, Ofer and Gross, Eilam",
    title = "{Estimating the significance of a signal in a multi-dimensional search}",
    eprint = "1105.4355",
    archivePrefix = "arXiv",
    primaryClass = "astro-ph.IM",
    doi = "10.1016/j.astropartphys.2011.08.005",
    journal = "Astropart. Phys.",
    volume = "35",
    pages = "230--234",
    year = "2011"
}

@article{grosso2025multiple,
  title={Multiple testing for signal-agnostic searches for new physics with machine learning},
  author={Grosso, Gaia and Letizia, Marco},
  journal={The European Physical Journal C},
  volume={85},
  number={1},
  pages={4},
  year={2025},
  publisher={Springer}
}

@article{hein2025,
  author        = {Hein, Marie and Nachman, Benjamin and Shih, David},
  title         = {Look everywhere effects in anomaly detection},
  year          = {2025},
  eprint        = {2512.13787},
  archivePrefix = {arXiv},
  primaryClass  = {hep-ph}
}

@article{dAgnolo:2021aun,
    author = "d'Agnolo, Raffaele Tito and Grosso, Gaia and Pierini, Maurizio and Wulzer, Andrea and Zanetti, Marco",
    title = "{Learning new physics from an imperfect machine}",
    eprint = "2111.13633",
    archivePrefix = "arXiv",
    primaryClass = "hep-ph",
    doi = "10.1140/epjc/s10052-022-10226-y",
    journal = "Eur. Phys. J. C",
    volume = "82",
    number = "3",
    pages = "275",
    year = "2022"
}

@misc{tong2026covariant,
  author       = {Tong, Shelley and Grosso, Gaia and Harris, Philip},
  title        = {Covariant Contrastive Learning for Uncertainty-Aware Anomaly Detection},
  howpublished = {PAI 2026 Conference (non-archival)},
  year         = {2026},
  url          = {https://openreview.net/forum?id=W2Q1F9EIdK}
}

@article{kuchibhotla2022postselection,
  author  = {Kuchibhotla, Arun K. and Kolassa, John E. and Kuffner, Todd A.},
  title   = {Post-Selection Inference},
  journal = {Annual Review of Statistics and Its Application},
  volume  = {9},
  pages   = {505--527},
  year    = {2022},
  doi     = {10.1146/annurev-statistics-100421-044639},
  url     = {https://doi.org/10.1146/annurev-statistics-100421-044639}
}

@article{skilling2006nested,
  title={Nested sampling for general Bayesian computation},
  author={Skilling, John},
  year={2006}
}

@inproceedings{cohen2016group,
  title={Group equivariant convolutional networks},
  author={Cohen, Taco and Welling, Max},
  booktitle={International conference on machine learning},
  pages={2990--2999},
  year={2016},
  organization={PMLR}
}

@article{dangovski2021equivariant,
  title={Equivariant contrastive learning},
  author={Dangovski, Rumen and Jing, Li and Loh, Charlotte and Han, Seungwook and Srivastava, Akash and Cheung, Brian and Agrawal, Pulkit and Solja{\v{c}}i{\'c}, Marin},
  journal={arXiv preprint arXiv:2111.00899},
  year={2021}
}

@article{metropolis1953equation,
  title={Equation of state calculations by fast computing machines},
  author={Metropolis, Nicholas and Rosenbluth, Arianna W and Rosenbluth, Marshall N and Teller, Augusta H and Teller, Edward},
  journal={The journal of chemical physics},
  volume={21},
  number={6},
  pages={1087--1092},
  year={1953},
  publisher={American Institute of Physics}
}

@article{hastings1970monte,
  title={Monte Carlo sampling methods using Markov chains and their applications},
  author={Hastings, W Keith},
  year={1970},
  publisher={Oxford University Press}
}

@article{bommasani2021opportunities,
  title={On the opportunities and risks of foundation models},
  author={Bommasani, Rishi and Hudson, Drew A and Adeli, Ehsan and Altman, Russ and Arora, Simran and von Arx, Sydney and Bernstein, Michael S and Bohg, Jeannette and Bosselut, Antoine and Brunskill, Emma and others},
  journal={arXiv preprint arXiv:2108.07258},
  year={2021}
}

@article{yang2023diffusion,
  title={Diffusion models: A comprehensive survey of methods and applications},
  author={Yang, Ling and Zhang, Zhilong and Song, Yang and Hong, Shenda and Xu, Runsheng and Zhao, Yue and Zhang, Wentao and Cui, Bin and Yang, Ming-Hsuan},
  journal={ACM computing surveys},
  volume={56},
  number={4},
  pages={1--39},
  year={2023},
  publisher={ACM New York, NY, USA}
}

@article{kaplan2020scaling,
  title={Scaling laws for neural language models},
  author={Kaplan, Jared and McCandlish, Sam and Henighan, Tom and Brown, Tom B and Chess, Benjamin and Child, Rewon and Gray, Scott and Radford, Alec and Wu, Jeffrey and Amodei, Dario},
  journal={arXiv preprint arXiv:2001.08361},
  year={2020}
}

@article{hoffmann2022training,
  title={Training compute-optimal large language models},
  author={Hoffmann, Jordan and Borgeaud, Sebastian and Mensch, Arthur and Buchatskaya, Elena and Cai, Trevor and Rutherford, Eliza and Casas, DDL and Hendricks, Lisa Anne and Welbl, Johannes and Clark, Aidan and others},
  journal={arXiv preprint arXiv:2203.15556},
  volume={10},
  year={2022}
}

@article{bahri2020statistical,
  title={Statistical mechanics of deep learning},
  author={Bahri, Yasaman and Kadmon, Jonathan and Pennington, Jeffrey and Schoenholz, Sam S and Sohl-Dickstein, Jascha and Ganguli, Surya},
  journal={Annual review of condensed matter physics},
  volume={11},
  number={1},
  pages={501--528},
  year={2020},
  publisher={Annual Reviews}
}

@article{
bahri_pnas,
author = {Yasaman Bahri  and Ethan Dyer  and Jared Kaplan  and Jaehoon Lee  and Utkarsh Sharma },
title = {Explaining neural scaling laws},
journal = {Proceedings of the National Academy of Sciences},
volume = {121},
number = {27},
pages = {e2311878121},
year = {2024},
doi = {10.1073/pnas.2311878121},
URL = {https://www.pnas.org/doi/abs/10.1073/pnas.2311878121},
eprint = {https://www.pnas.org/doi/pdf/10.1073/pnas.2311878121}}


\end{document}